\documentclass[iop,twocolumn,numberedappendix]{emulateapj}
\bibpunct[; ]{(}{)}{;}{a}{}{;}
\usepackage{apjfonts}

\usepackage{graphicx} 

\newcommand{\Msolar}{M${_\odot}$\,}

\shorttitle{Photometric determination of the mass accretion rates of
pre-main sequence stars in NGC\,346}
\shortauthors{De Marchi et al.}

\slugcomment{Accepted for publication in ``The Astrophysical Journal''}

\begin{document}

\title{Photometric determination of the mass accretion rates of pre-main
sequence stars. II. \\ NGC\,346 in the Small Magellanic
Cloud\,\altaffilmark{*}}


\author{
Guido De Marchi,\altaffilmark{1}
Nino Panagia,\altaffilmark{2,3,4} 
Martino Romaniello,\altaffilmark{5}
Elena Sabbi,\altaffilmark{2}
Marco Sirianni,\altaffilmark{1}\\
Pier Giorgio Prada Moroni\altaffilmark{6,7} and
Scilla Degl'Innocenti\altaffilmark{6,7}
}

\altaffiltext{1}{European Space Agency, Space Science Department,
Keplerlaan 1, 2200 AG Noordwijk, Netherlands; gdemarchi@rssd.esa.int}

\altaffiltext{2}{Space Telescope Science Institute, 3700 San Martin Drive, 
Baltimore, MD 21218, USA, panagia@stsci.edu}
 
\altaffiltext{3}{INAF--CT, Osservatorio Astrofisico di Catania, Via S.
Sofia  78, 95123 Catania, Italy}

\altaffiltext{4}{Supernova Limited, OYV \#131, Northsound Rd., Virgin
Gorda,  British Virgin Islands}

\altaffiltext{5}{European Southern Observatory, Karl--Schwarzschild-Str.
2, 85748 Garching, Germany}

\altaffiltext{6}{Dipartimento di Fisica ``Enrico Fermi'', Universit\`a
di Pisa, Largo Pontecorvo 3, 56127 Pisa, Italy}

\altaffiltext{7}{INFN -- Sezione di Pisa, Largo Pontecorvo 3, 56127
Pisa, Italy}

\altaffiltext{{$\star$}}{Based on observations with the NASA/ESA
{\it Hubble Space Telescope}, obtained at the Space Telescope Science
Institute, which is operated by AURA, Inc., under NASA contract
NAS5-26555}

\begin{abstract}  

We have studied the properties of the stellar populations in the field
of the NGC\,346 cluster in the Small Magellanic Cloud, using a novel 
self-consistent method that allows us to reliably identify pre-main 
sequence (PMS) objects actively undergoing mass accretion, regardless of
their age. The method does not require spectroscopy and combines
broad-band $V$ and $I$ photometry with narrow-band $H\alpha$ imaging to
identify all stars with excess H$\alpha$ emission and derive the
accretion luminosity $L_{\rm acc}$ and mass accretion rate $\dot M_{\rm
acc}$ for all of them. The application of this method to existing
HST/ACS photometry of the NGC\,346 field has allowed us to identify and
study 680 bona-fide PMS stars with masses from $\sim 0.4$\,\Msolar to
$\sim 4$\,\Msolar and ages in the range from $\sim 1$\,Myr to $\sim
30$\,Myr. {Previous investigations of this region, based on the
same data, had identified young ($\sim 3$\,Myr old) candidate PMS stars
on the basis of their broad-band colours. In this study we show that
there are at least two, almost equally numerous, young populations with 
distinct ages of respectively $\sim 1$ and $\sim 20$ Myr. We provide for 
all of them accurate physical parameters.} 

We take advantage of the unprecedented size of our PMS sample and of its
spread in mass and age to study the evolution of the mass accretion rate
as a function of stellar parameters. We find that, regardless of stellar
mass, the mass accretion rate decreases with roughly the square root of
the age, or about three times slower than predicted by current models of
viscous disc evolution, and that more massive stars have systematically
higher mass accretion rate in proportion to their mass. A multivariate
linear regression fit reveals that $\log \dot M_{\rm acc} \simeq -0.6
\log t + \log m + c$, where $t$ is the age of the star, $m$ its mass and
$c$ {a quantity that is higher at lower metallicity. This result is
consistent with measurements of the mass accretion rate in the 30 Dor 
region and in the Milky Way and suggests that longer duration for mass
accretion could be related to lower metallicity}. The high mass
accretion rates that we find suggest that a considerable amount of mass
is accreted during the PMS phase, of order {$\sim 0.2$\,\Msolar  or
possibly $\sim 20\,\%$ of the final mass for stars with mass $m <
1$\,\Msolar if their discs are eroded by 20\,Myr, i.e. before they
reach the main sequence. Therefore,} PMS evolutionary models that do not
account for this effect will systematically underestimate the true age
when compared with the observations.

\end{abstract}


\section{Introduction}
\label{intro}

NGC\,346 is presently the region of most intense star formation in the
entire Small Magellanic Cloud (SMC). It contains over 30 O type stars
(Massey, Parker \& Garmany 1989; Evans et al. 2006) that ionise the
N\,66 nebula, the largest HII region in the SMC (Henize 1956). The
presence of these very young ($\sim 1-3$\,Myr) massive stars in NGC\,346
has long been established with ground-based spectroscopy (e.g. Walborn
1978; Walborn \& Blades 1986; Niemela et al. 1986; Massey et al. 1989),
but more recently also very young stars of low mass have been detected.
Observations with the Hubble Space Telescope and Spitzer Space Telescope
have revealed respectively a multitude of pre-main sequence (PMS) star
candidates down to the subsolar mass (Nota et al. 2006; Sabbi et al.
2007; Gouliermis et al. 2007; Hennekemper et al. 2008) and a large
number of candidate young stellar objects (YSO) with masses as low as
$1.5$\,\Msolar (Bolatto et al. 2007; Simon et al. 2007).

What makes NGC\,346 and the surrounding regions particularly interesting
is that they allow us to study the properties of star formation in a
{nearby galaxy with a metallicity similar to those in place in the
high redshift universe at $z>2$. The distance modulus for the SMC is
$18.92 \pm 0.03$ corresponding to $\sim 61$\,kpc (Hilditch et al. 2005;
Keller \& Wood 2006), whereas the currently accepted values for its
metallicity range from  $\sim 1/5$ to $\sim 1/8$\,Z$_\odot$ (see Russell
\& Dopita 1992; Rolleston et al. 1999; Lee et al. 2005; Perez--Montero
\& Diaz 2005)}. Moreover, besides this recent burst, previous star
formation episodes have been detected in this area. Heap et al. (2006)
and Mokiem et al. (2006) reported the presence of massive stars with an
age of $\sim 5$\,Myr, while Massey et al. (1989) discovered five red
supergiants and two B type supergiants that form a spatially distinct
subgroup located about $2\farcm6$ SW of the centre of NGC\,346 and that
have an  estimated age of $\sim 12$\,Myr. More recently, Sabbi et al.
(2007)  detected a small star cluster, located about $2\farcm1$ or $\sim
37$\,pc  NE of the centre whose colour--magnitude diagram (CMD) is
compatible with an age of $15 \pm 2.5$\,Myr. Finally, older populations
are also present: Sabbi et al. (2007) find evidence for star formation
in this region dating as far back as $\sim 10$\,Gyr with a moderate
enhancement $\sim 150$\,Myr ago. In addition, an intermediate-age
cluster, BS\,90, is also present in the field, with an age of $4.3\pm
0.1$\,Gyr.

{It is not easy to establish whether there is any relationship
between the very old generations in this field and the more recent star
formation episodes. However, it is interesting to investigate whether
the $\sim 12$\,Myr old massive stars of Massey et al. (1989) and the $15
\pm 2.5$\,Myr old small cluster of Sabbi et al. (2007), although
spatially unrelated to one another, could be the precursors of the 30
young massive O type stars and of the recently detected PMS stars and
YSO in this region. It is conceivable that a large number of these $\sim
15$\,Myr old objects are actually present in NGC\,346, many more than
those known so far, but that they could belong to a more diffuse
population spread throughout the region because of appreciable velocity
dispersion motions and, therefore, might not be easy to identify. If the
members of such a $\sim 15$\,Myr old population could be detected,
including those of low mass, their spatial distribution and physical
properties would allow us to understand whether and how they may have
triggered the most recent bursts, as theories of sequential star
formation suggest (e.g. Elmegreen \& Lada 1977).}

The task of detecting a $\sim 15$\,Myr old population in a region
heavily contaminated by older field stars and younger objects might
appear daunting. Hennekemper et al. (2008) attempted to study the
presence of an age spread in the PMS population of NGC\,346 through the
analysis of the CMD. They showed that the observations are in principle
compatible with an upper limit of $\sim 10$\,Myr to the age, but they
concluded that differential reddening and unresolved binaries could be
at the origin of the observed colour spread.

{Actually, if these objects have not yet completed their pre-main
sequence (PMS) phase, it should be possible to detect them through the
distinctive excess emission features in their spectra that originate
from the accretion process, particularly in the UV continuum and in
recombination lines such as H$\alpha$, Pa$\beta$ and Br$\gamma$ (e.g.
Calvet et al. 2000). Although in nearby star forming regions the
fraction of PMS stars with discs appears to decline quite rapidly during
the first 10\,Myr (e.g. Haisch, Lada \& Lada 2001; Fedele et al. 2010),
there is evidence that sustained mass accretion is present also in
objects older than this age, particularly above 1\,\Msolar. This is the
case for instance of Tr\,37 where Sicilia--Aguilar et al. (2006) find
G-type dwarfs still accreting at ages in excess of 10\,Myr. In the more
distant ($\sim 7$\,kpc) and massive Galactic star forming region
NGC\,3603 Beccari et al. (2010) recently found a conspicuous number of
PMS stars older than 10\,Myr still undergoing mass accretion. In even
more distant and dense stellar fields, where spectroscopy of individual
stars is limited to the brightest members, Romaniello (1998), Panagia et
al. (2000) and Romaniello, Robberto \& Panagia (2004) have shown that
objects undergoing active mass accretion can be efficiently detected
with accurate multi-colour photometry} and have identified in the
regions around SN\,1987A about 500 PMS stars with H$\alpha$ equivalent
width in excess of 8\,\AA\ and with an age of $12 \pm 2$\,Myr. More
recently, De Marchi, Panagia \& Romaniello (2010, hereafter Paper\,I)
showed that through a suitable combination of broad- and narrow-band
photometry it is also possible to derive the mass accretion rate of
these objects, with an accuracy comparable to that allowed by
spectroscopy.

In this work, we build on the method developed in Paper\,I to securely
identify bona-fide PMS objects in dense stellar fields and apply it to
the high-quality HST photometry of the NGC\,346 region (Sabbi et al. 
2007). Our goal is to measure the physical properties of these objects
and study how they depend on age, in order to understand how star
formation has proceeded in this area over the past $\sim 30$\,Myr. The
paper is organised as follows: in Section\,\ref{obser} we briefly
describe the observations and address how we correct for differential
reddening, whereas Section\,\ref{pmsst} is devoted to the identification
of PMS stars via their H$\alpha$ excess emission and on the
determination of their H$\alpha$ luminosity. In Section\,\ref{physi} we
derive the other important physical parameters for these objects,
including age, mass and mass accretion rate, while in
Section\,\ref{evolu} we study how the latter evolves in time, how it
could affect stellar evolution, and we compare the results to those for
similar objects in the Milky Way and Large Magellanic Cloud. A summary
of the most important conclusions of the paper is offered in
Section\,\ref{summa}.

\begin{figure}
\centering
\resizebox{\hsize}{!}{\includegraphics[bb=50 20 565 486,width=16cm]
{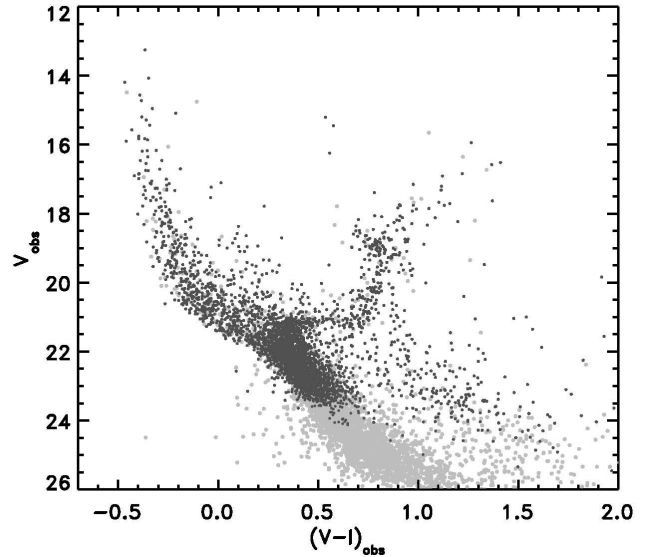}}
\caption{The observed colour--magnitude diagram of a field of $\sim 200
\arcsec$ on a side containing the cluster NGC\,346 in the Small
Magellanic Cloud, as observed with the Advanced Camera for Surveys on
board the Hubble Space Telescope by Nota et al. (2006). The diagram,
based on the  photometric study of Sabbi et al. (2007), contains  and a
total of about 53\,550  stars detected at the $> 3\,\sigma$ level in $V$
and $I$, of which $18\,764$ have a combined magnitude uncertainty in the
$V$, $I$ and H$\alpha$ bands of less than $0.1$\,mag (see
Equation\,\ref{eq1}).} 
\label{fig1}
\end{figure}

\section{The observations}
\label{obser}

The observations used in this work were collected with the Wide Field
Channel of the Advanced Camera for Surveys (Ford et al. 2003) on board 
the Hubble Space Telescope in 2004 July as part of proposal GO--10248
(with Antonella Nota as principal investigator). The data and their
reduction were extensively described in Sabbi et al. (2007), to which we
refer the reader for details. In summary, the observations of interest
here cover an area of $\sim 200 \arcsec$ on a side centered on the
cluster NGC\,346 in the Small Magellanic Cloud (SMC), at RA$=00^{\rm h}
59^{\rm m} 5.2^{\rm s}$,  DEC=$-72^\circ 10^\prime 28\arcsec$ (J2000),
itself immersed in N\,66, the largest and brightest HII region in the
SMC.

{The data comprise both deep and short exposures in the broad bands
F555W (hereafter $V$) and F814W (hereafter $I$), as well as observations
in the  80\,\AA\ wide band F658N (hereafter H$\alpha$), which were not
described in Sabbi et al. (2007).} The total combined exposure times in
each band are, respectively, 4\,148\,s, 4\,124\,s and 1\,542\,s. {The
data reduction {in the $V$ and $I$ bands} was carried out by Sabbi et
al. (2007) using a combination of PSF fitting and aperture photometry
and their final catalogue, in the central $200\arcsec \times 200\arcsec$
of interest here, includes about 53\,550 stars reliably detected (at the
$> 3\,\sigma$ level) in  $V$ and $I$. As regards the H$\alpha$ band, we
have adopted the same combination of PSF fitting and aperture photometry
to search for and measure all objects with a signal in the central pixel
exceeding the local background level by at least 3  times the standard
deviation of the latter (i.e. $> 3\,\sigma$ detection). Of all the stars
detected in the $V$ and $I$ bands, about 19\,800 satisfy this condition
also in the H$\alpha$ band.}

\begin{figure*}
\centering
\resizebox{\hsize}{!}{\rotatebox{90}{\includegraphics[width=16cm]
{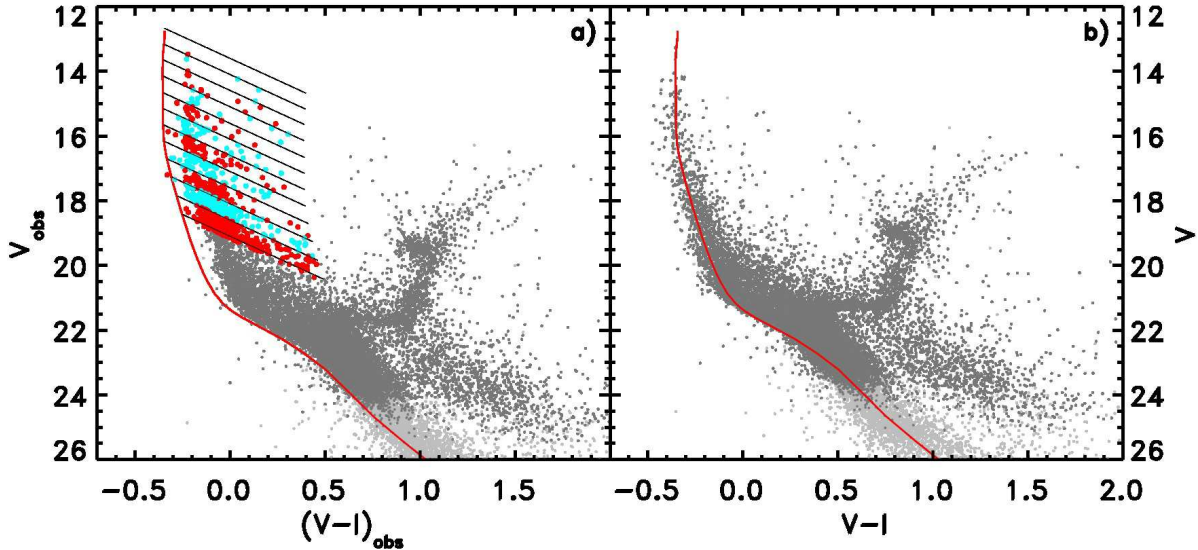}}}
\caption{Panel {\em a}): Observed CMD as in Figure\,\ref{fig1}, with
overplotted the isochrone of Marigo et al. (2008) for an age of 4\,Myr,
metallicity $Z=0.002$ and distance modulus $(m-M)_0=18.92 \pm 0.03$. The
slanted thin solid lines show the reddening vectors, {for $A_V=2$,}
according to the extinction law of Scuderi et al. (1996). They are drawn
every $0.5$ mag for stars brighter than $V_{\rm obs} \simeq 18.5$ and
are used to identify likely stars in the upper MS that have been
displaced by reddening (shown as thick dots). Panel {\em b}): the same
CMD after correction for differential reddening, using as a reference
for each object the average extinction value of the twenty nearest upper
MS stars. Note that both the upper MS and RC are tight.} 
\label{fig2}
\end{figure*}

We show in Figure\,\ref{fig1} the resulting colour--magnitude diagram
(CMD) of all the stars in the selected field as light grey dots and
with darker symbols stars whose mean error $\delta_3$ in
the three bands does not exceed $0.1$\,mag, where

\begin{equation} 
\delta_3 = \sqrt{\frac{\delta^2_{555}+\delta^2_{658}+\delta^2_{814}}{3}}
\label{eq1} 
\end{equation} 

\noindent 
and $\delta_{555}$, $\delta_{658}$ and $\delta_{814}$ are the
photometric uncertainties in each individual band. A total of $18\,764$ 
stars satisfy this condition, with the most  stringent constraint being
set by the uncertainty in the H$\alpha$ band.

The CMD, as already discussed by Nota et al. (2006), Gouliermis et al. 
(2006; 2007) and Sabbi et al. (2007), reveals the presence in this field
of both young and old populations. The former includes the massive stars
on the pronounced upper main sequence (MS) and for the candidate PMS
objects to the right of and brighter than the lower MS, while the latter
accounts for the lower MS and for the sub-giant and red giant branches
(SGB, RGB). In general, however, it is not possible to classify an
object as being young or old just on the basis of its position on the
CMD, not only because there are regions such as the lower MS that are
common to both populations, but also because there is evidence for
variable reddening in the field, as Sabbi et al. (2007) already pointed
out and as one can see from the conspicuous broadening of the upper MS
in Figure\,\ref{fig1}. 

Differential reddening makes it particularly difficult to understand the
true nature of the large group of objects below the SGB and redder than
the MS. While their colours and magnitudes, as originally pointed out by
Nota et al. (2006), are compatible with those of young PMS stars,
depending on the amount of reddening and on the specific extinction law
for this field, some of them may in reality be reddened MS or SGB stars.
Furthermore, even if all these objects were PMS stars, without
accounting for differential reddening one could not properly assign to
them a mass and age through the comparison with PMS evolutionary tracks
(see e.g. Hennekemper et al. 2008). {It is  therefore important to
address the issue of differential reddening and to correct the
photometry accordingly, to the extent possible. We do so in the
following section.}

\subsection{Reddening correction}
\label{redde}

Sabbi et al. (2007) concluded that the young population in this field,
corresponding to the upper MS, is compatible with an age of $4 \pm
1$\,Myr. Therefore, one can determine the reddening towards each of
those stars if one assumes that they are all MS objects and one looks at
their colour and magnitude displacement with respect to a theoretical
isochrone for an age of 4\,Myr, a metallicity of $Z=0.002$ and a
distance modulus of $18.92 \pm 0.03$ (see Introduction). Hennekemper et
al. (2008) followed a similar approach to derive an extinction map for
this region, albeit with a set of isochrones for a higher metallicity
($Z=0.004$). We note here that although the metallicity that we adopt,
$Z=0.002$, is at the lower end of the currently accepted values for the
SMC {(see Introduction)}, it appears better suited to describe the
properties of young PMS stars, as we will show in Section\,\ref{physi}.

As for the extinction law in this field, we decided to use the one
derived by Scuderi et al. (1996) for the field of SN\,1987A in the Large
Magellanic Cloud, rather than the Galactic law as adopted by Hennekemper
et al. (2008), in light of the more similar metallicity. {The total
extinction values for the specific HST filters used here were derived by
averaging over the entire bandpass the dust attenuation ($10^{-0.4
A_\lambda}$), as explained in Romaniello et al. (2002)}. The solid lines
cutting through the upper MS in Figure\,\ref{fig2}a correspond to the
reddening vectors drawn from a 4\,Myr old isochrone for $Z=0.002$
(Marigo et al. 2008), in steps of $0.5$ mag for stars brighter than
$V_{\rm obs} \simeq 18.5$. {The length of the lines corresponds to
$A_V=2$}. A total of about 600 stars selected in this way and bluer than
$V-I=0.3$ are marked with thicker symbols in Figure\,\ref{fig2}a. The
corresponding reddening distribution ranges from $E(V-I)=0.1$ to
$E(V-I)=0.25$ (respectively 17\,\% and 83\,\% limits {corresponding
to $A_V$ values of $0.27$ and $0.67$}), with a median value of
$E(V-I)=0.16$ or $A_V=0.43$, indicating that a considerable spread in
reddening is present in this field. 


{We have used the reddening values towards each of the selected upper MS
stars to derive a reddening correction for all other objects in their
vicinities. The analysis of the H$\alpha$ image indicates that the
typical scale of the brightness fluctuations is of order $5\arcsec$
radius. We have therefore assumed that stars projected within this
region have similar extinction, even though we have no information on
the depth of their distribution nor on the uniformity of the absorbing
material. In order to validate the procedure, we first applied the
reddening correction to the upper MS stars themselves, using for each
object the average of the extinction values of at most 20 other upper MS
stars within $5\arcsec$ of their location. We took the straight average
{without weighting them} for the distance from the object, since we
have no information on the distribution of the stars along the line of
sight.  We found that the average and  standard deviation of the
reddening distribution derived with the nearest neighbours are in
excellent agreement with those measured directly, namely $<A_V>=0.51 \pm
0.19$. This has convinced us that  the average reddening value measured
in a small area surrounding each star can be used to correct for
extinction. Objects with no reference upper MS stars within $5\arcsec$
were assigned the average reddening value of all reference stars
($A_V=0.51$). }

The upper MS corrected in this way appears considerably tighter in the
CMD than the raw data (see Figure\,\ref{fig2}b) and this convinced us
that the method, albeit statistical in nature, can be applied to the
entire data set. We also experimented with a different number of
neighbours, namely 5, 10, 20 and 50. We found that, by using the 20
nearest upper MS stars, {the red clump (RC) feature in the red giant
branch of old ($\sim 1$\,Gyr)} in the CMD at $V\simeq 19$ appeared the
tightest. Since our goal is to correct for reddening all candidate PMS
stars, which are typically red, this choice seemed the most appropriate.
The de-reddened CMD obtained in this way is shown in
Figure\,\ref{fig2}b.

\vspace*{0.5cm}
\section{Identifying the pre-main sequence stars}
\label{pmsst}

One of the characteristic signatures of the accretion process on to PMS
stars is the presence of excess emission in H$\alpha$ (see e.g. Calvet
et al. 2000). Recently, we showed in Paper\,I that PMS stars undergoing
active accretion can be reliably identified, and their H$\alpha$
luminosity measured, by using a combination of broad-band ($V, I$) and
narrow-band (H$\alpha$) photometry, with no need for spectroscopy. This
way of identifying PMS stars is more accurate and reliable than the
simple classification based on the position of the objects in the
Hertzsprung--Russell diagram, i.e. stars placed well above the main
sequence (e.g. Gilmozzi et al. 1994; Hunter et al. 1995; see in
particular Nota et al. 2006 and Gouliermis et al. 2007 for the specific
case of NGC\,346). The true advantages of the method presented in
Paper\,I are that: {\em (i)} it allows us to discriminate between
bona-fide PMS stars and interlopers that occupy the same region of the
CMD; {\em (ii)} it provides a secure detection of relatively old PMS
stars, already close to their MS, whose colours and magnitudes would
otherwise be indistinguishable from those of MS stars; {\em (iii)} by
measuring the $H\alpha$ luminosity $L(H\alpha)$ of these PMS stars we
can obtain their mass accretion rates $\dot M_{\rm acc}$ from
$L(H\alpha)$ and the stellar parameters (mass and radius);  {\em (iv)}
finally, this approach is more practical, efficient and economical than
slitless spectroscopy for studying PMS stars in dense star forming
regions, although it cannot provide constraints for magnetospheric
accretion models that require information on the profile of the
H$\alpha$ emission line (e.g. Muzerolle et al. 2001).

As explained in detail in Paper\,I, the method hinges on the simple 
consideration that, at any time, the largest majority of stars of a
given effective temperature $T_{\rm eff}$ in a stellar population will
have no excess H$\alpha$ emission. Therefore, for stars of that
effective temperature, the {\em median} value of any colour index
involving H$\alpha$  (e.g. $V-H\alpha$) effectively defines a spectral
reference template with respect to which the H$\alpha$ colour excess
should be sought.  

\begin{figure}
\centering
\resizebox{\hsize}{!}{\includegraphics[bb=50 20 565 486,width=16cm]
{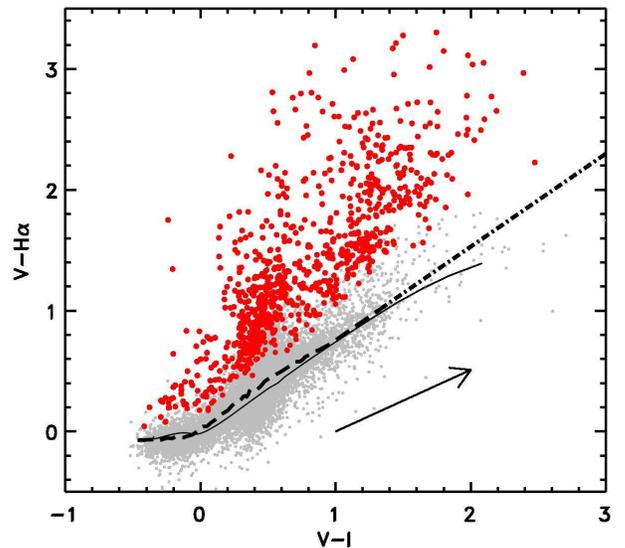}}
\caption{Colour--colour diagram of the selected 18\,764 stars with
$\delta_3 \le 0.1$\,mag. The dashed line represents the running median  
$V-H\alpha$ colour, obtained with a box-car size of 100 points, whereas 
the thin solid line shows the model atmospheres of Bessell et al.
(1998). The arrow in the figure corresponds to the reddening vector for
$A_V=2.7$ or $E(V-I)=1$, taken as an example, for the specific bands
used in this study. A total of 791 objects with a $V-H\alpha$ excess
larger than $4\,\sigma$ are indicated with large dots (in red in the
online version).} 
\label{fig3}
\end{figure}

In Figure\,\ref{fig3} we show with small dots the $V-H\alpha$ colour as
a function of $V-I$ for all 18\,764 stars that have a combined
photometric uncertainty $\delta_3 \le 0.1$\,mag (see
Equation\,\ref{eq1}). The value of $\delta_3$ is dominated by the
uncertainty on the H$\alpha$ magnitude, while the median value of the
uncertainty in the other two bands is $\delta_V=0.006$ and
$\delta_I=0.008$, respectively.  

The thick dashed line in Figure\,\ref{fig3} represents the {median}
$V-H\alpha$ colour obtained as the running median with a box-car size of
100 points. The thin solid line shows the colours in these filters for
the model atmospheres of Bessell et al. (1998). As discussed in
Paper\,I, the $< 0.1$\,mag discrepancy between models and observations
around $V-I\simeq 0.5$ is most likely due to the coarse spectral
sampling of the H$\alpha$ line in the models. Where the density of
observed points decreases considerably, at $V-I > 1.2$, we  reduced the
box-car size to 10 points and the result is shown by the  thick
dot-dashed line (which is extrapolated for $V-I>2$). A total of 891
objects have a $V-H\alpha$ index exceeding that of the  reference
template {(i.e. the photospheric colour)} at the same $V-I$ colour
by more than four times the uncertainty on their $V-H\alpha$ values. 

{The arrow in the figure corresponds to the reddening vector for
$A_V=2.7$ or $E(V-I)=1$ for the specific bands used in this study,
derived from the extinction law of Scuderi et al. (1996). The fact that
the arrow is almost parallel to the reference template implies that even
relatively large uncertainties in the reddening correction would not
significantly affect the selection of stars with H$\alpha$ excess. On
the other hand, it is possible that for some objects surrounded by non
homogeneous nebular emission the H$\alpha$ excess that we measure is
spurious. We explain in Section\,\ref{nebul} how we have removed these
objects from our list.

\subsection{Nebular contamination}
\label{nebul}

The effects of nebular contamination on the detection of stars with
H$\alpha$ excess emission are addressed in detail in Paper\,I (and in
Beccari et al. 2010), to which we refer the reader. We show there that
gas emission acts like a ``pedestal,'' or an additional source of
background. As such, it does not affect our photometry because it
increases the measured flux in the same way both inside the central
aperture and inside the background annulus used for the determination of
the sky level. Therefore the gas emission is automatically removed in
the sky subtraction procedure. If the diffuse emission is strong
compared to the intensity of the objects, the  uncertainty on the flux
of the star will be larger, but this is fully taken into account in our
photometry. 

A possible complication, not discussed in Paper\,I, arises when the
nebular emission is not homogeneous across the field. For example, when
a filamentary structure in the nebular gas overlaps with a star, the
magnitude of the star can be affected in different ways. If the filament
is wider than the diameter of the annulus that we use to measure the
local background, this only results in a higher photometric uncertainty,
because we fall in the case explained above for diffuse emission. The
situation is more complex when the filament only partly covers the
background annulus (which in our case  extends from 3 to 5 pixel radius
around the object) and/or the central star (for which an aperture of 2
pixel radius is used), since in those cases the signal from the star
and/or the background could be incorrectly determined.

We, therefore, inspected the H$\alpha$ image around the location of all 
stars with an apparent H$\alpha$ excess. In order to enhance the
contrast between neighbouring pixels in the image, we applied an
unsharp-masking filter (e.g. Malin 1977) to the H$\alpha$ frame, with a
smoothing radius of 10 pixel meant to subtract the low-frequency signal,
i.e. structures in the image with a characteristic scale larger than the
size of our background annulus. Unsharp masking makes structures and
filaments in the nebular gas easier to see than in the direct image. In
this way we compared the structures and filaments with the positions of
the stars with Halpha excess emission (see Beccari et al. 2010 for an
application of this method to the case of NGC\,3603). We adopted a very
conservative approach and excluded from our PMS candidate list all stars
whose photometric background annulus is partially covered by a gas
filament. In total, about 70 stars were excluded in this way, or less
than 10\,\% of the original population (an additional 30 stars were
excluded because they are closer than 10 pixels to the image edges in
the Halpha frame, although their magnitudes are well defined in the
other bands). }

A total of 791 objects are left after this conservative selection and
they are indicated with large dots in Figure\,\ref{fig3} (shown in red
in the on-line version) and, in light of their $V-H\alpha$ excess at the
$4\,\sigma$ level, they are excellent PMS candidates. As we shall see
later, most of them are indeed bona-fide PMS stars.

\subsection{H$\alpha$ luminosity}

Following the detailed procedure outlined in Paper\,I, we can derive the
H$\alpha$ emission line luminosity $L(H\alpha)$ for these stars from the
$\Delta H\alpha$ colour corresponding to the excess emission, namely:

\begin{equation}
\Delta H\alpha =  (V-H\alpha)^{\rm obs} - (V-H\alpha)^{\rm ref}
\label{eq2}
\end{equation}

\noindent   
where the superscript {\em obs} refers to the observations and {\em ref}
to the reference template. $L(H\alpha)$ is then obtained from $\Delta
H\alpha$, from the photometric zero point and absolute sensitivity of
the instrumental set-up and from the distance to the sources. {The
assumed distance modulus is $(m-M)_0=18.92$ (see Introduction)}, whereas
the photometric  properties of the instrument are as listed in the ACS
Instrument  Handbook (Maybhate et al. 2010).
The derived median $H\alpha$ luminosity of the 791 objects with
H$\alpha$ excess at the $4\,\sigma$ level is $2.7 \times 10^{31}$\,erg/s
or $\sim 10^{-2}$\,L$_\odot$. This value is similar to that of a group
of 189 stars with H$\alpha$ excess emission in the field of SN\,1987A
that were studied in Paper\,I with the same method applied to HST Wide
Field Planetary Camera\,2 data. The uncertainty on $L(H\alpha)$ is
typically $\sim 12$\,\% and is completely dominated by the statistical
uncertainty on the H$\alpha$ photometry (10\,\%), while the distance
accounts for 5\,\% (Hilditch et al. 2005; Keller \& Wood 2006) and the
absolute sensitivity of the instrumental setup for another 3\,\%. 

Owing to the rather wide passband of the F658N filter, $\Delta H\alpha$
includes a small contribution from the [NII] emission features at 
$6\,548$\,\AA\, and $6\,584$\,\AA. Even though such a contribution is
small (see Paper\,I, {where the effect is explained in detail}), it
is a good practice to correct for it as it is a systematic effect. The
adopted average correction factor to the intensity for the ACS F658N
filter, {based on the spectra of classical T Tauri stars} is
$0.979$. 

\begin{figure}
\centering
\resizebox{\hsize}{!}{\includegraphics[bb=50 20 565 486,width=16cm]
{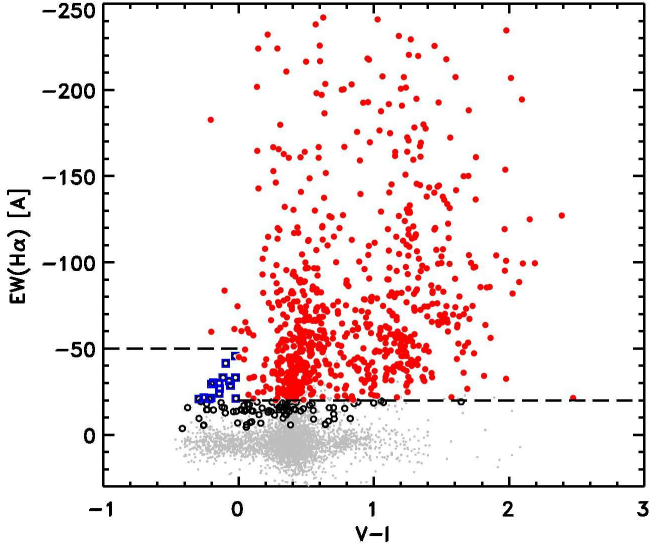}}
\caption{H$\alpha$ equivalent width of the stars in the field of
NGC\,346, as a function of their $V-I$ colour. Of the 791 stars with
H$\alpha$ excess at the $4\,\sigma$ level (circles and squares), a
total of 694 have $W_{\rm eq}(H\alpha) < -20$\,\AA\ (or $< -50$\,\AA\ for
stars hotter than 10\,000\,K). As such, we consider them as bona-fide
PMS stars and indicate them with thick dots (shown in red in the online
version). As customary, we use negative equivalent widths for emission
lines.} 
\label{fig4}
\end{figure}

\subsection{H$\alpha$ equivalent width}

Besides the luminosity of the H$\alpha$ emission line, also its
equivalent width $W_{\rm eq}(H\alpha)$ can be derived from our
photometry. To this aim, as shown in Paper\,I, we used the model
atmospheres of Bessell et al. (1998) to derive the magnitude
$H\alpha^c$ corresponding to the sole continuum in the H$\alpha$ band
from the $V$ and $I$ magnitudes. The corresponding relationships,
properly validated by comparison with spectro-photometric measurements,
are listed in the Appendix of Paper\,I and allow us to derive $W_{\rm
eq}(H\alpha)$ from the H$\alpha$ magnitude via the equation:

\begin{equation}
W_{\rm eq} (H\alpha) =  \mathrm{RW} \times [1-10^{-0.4 \times
(H\alpha-H\alpha^c)}]
\label{eq3}
\end{equation}

\noindent  where $\mathrm{RW}$ is the rectangular width of the filter
(similar in definition to the equivalent width of a line), which depends
on the characteristics of the filter and corresponds to $74.96$\,\AA\
for the F658N filter used here. The equivalent widths obtained in this
way are shown in Figure\,\ref{fig4} as a function of the $V-I$ colour.
The 791 stars with a $4\,\sigma$ H$\alpha$ excess are indicated with
thicker symbols (circles and squares). We take as bona-fide PMS stars
all those  with $W_{\rm eq}(H\alpha) < -20$\,\AA, indicated as thick
dots (in red in the online version), corresponding to a total of 694
stars. Note that, as customary, negative values of the equivalent width
indicate emission lines. Since at temperatures $T_{\rm eff} \ga
10\,000$\,K or colours $V-I \la 0$ the sample could be contaminated by
Be stars that are evolving off the MS, there we set a more stringent
condition on the equivalent width, namely $W_{\rm eq}(H\alpha) <
-50$\,\AA. This limit is suggested by a survey of the H$\alpha$
equivalent width of about 100 Be stars in the Galaxy (Cot\'e \& Waters
1987) in which only one star is found with $W_{\rm eq}(H\alpha) <
-50$\,\AA\ and the largest majority have values in the range from
$-4$\,\AA\ to $-30$\,\AA. In fact, as we discuss in
Section\,\ref{physi}, all the objects with $T_{\rm eff}> 10\,000$\,K and
an H$\alpha$ excess above $4\,\sigma$ are most likely PMS objects with
masses around 3\,\Msolar. 

{Although we do not have spectroscopic observations of our stars to 
directly confirm our photometrically inferred equivalent widths,
similar studies in the literature have demonstrated the success of this
approach.} White \& Ghez (2001) carried out a detailed study of 44 T
Tauri stars in binary systems in Taurus--Auriga, using both photometry
and spectroscopy with the HST. Their analysis shows that photometric and
spectroscopic measurements of the equivalent width of the H$\alpha$
emission line in these objects are in good agreement with one another.
More recently, Barentsen et al. (2011) carried out a H$\alpha$ survey of
T Tauri stars in IC\,1396, a region with considerable nebular emission.
Their photometric determination of $W_{\rm eq}(H\alpha)$ for 109 objects
in common with the spectroscopic survey of Sicilia--Aguilar et al.
(2005) also shows an excellent agreement, particularly for stars with
$W_{\rm eq}(H\alpha)< -10$\,\AA, which are the objects of interest in
our study.

\section{Physical parameters of the pre-main sequence stars}
\label{physi}

\begin{figure}
\centering
\resizebox{\hsize}{!}{\includegraphics[bb=50 20 565 486,width=16cm]
{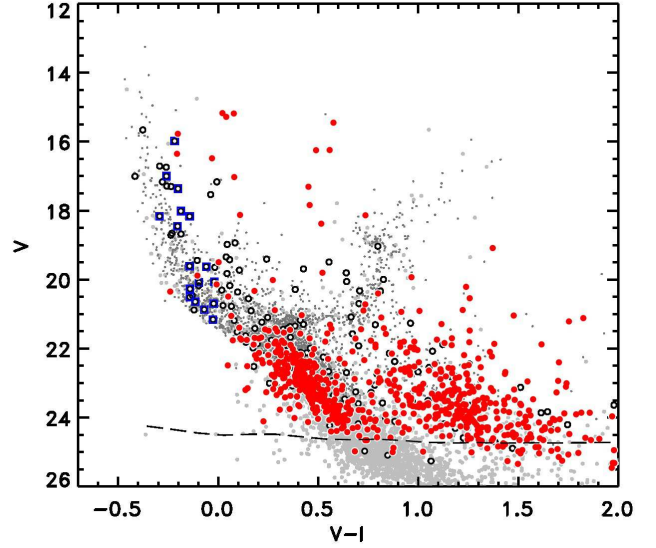}}
\caption{Colour--magnitude diagram of the field of NGC\,346 showing as
open circles the 791 PMS star candidates with excess emission in
H$\alpha$ at the $4\,\sigma$ level or higher. A total of 694 of them
(shown as filled circles, in red in the online version) also have  
$W_{\rm eq}(H\alpha) < -20$\,\AA\ or $< -50$\,\AA\ for stars hotter than
10\,000\,K. Squares correspond to objects with $W_{\rm eq}(H\alpha) >
-50$\,\AA\ that are potential Be stars. The dashed line corresponds to
the 50\,\% completeness limit of the photometry (see Sabbi et al.
2008).} 
\label{fig5}
\end{figure}

When the bona-fide PMS objects identified in the previous section are
placed on the CMD (see Figure\,\ref{fig5}), they clearly reveal two
distinct groups {made up of a similar number of objects}. One is
located above (i.e. brighter and redder than) the MS and thus is
indicative of a population of young PMS objects. This population was
originally noticed by Nota et al. (2006) and later studied by Gouliermis
et al. (2007) and Sabbi et al. (2007), although these earlier works
could only identify candidate young ($< 5$\,Myr) PMS stars since no
information was available to verify and characterise the PMS nature of
each individual object. This is the reason why the works above could not
identify the second group of bona-fide PMS {objects}, which overlap
in the CMD with older MS stars in the field of the SMC. Hennekemper et
al. (2008) {detected some of these objects using a preliminary
version of the colour transformations between $V$, $I$ and $R$ that we
developed for Paper\,I to search for stars with $R-H\alpha$ excess in
this field, but they did not realise what was the nature of these
objects.} As we show here below, these objects are older PMS stars with
an age in excess of $\sim 8$\,Myr, of the same type as those that we
discovered in the field around SN\,1987A in the LMC (see Paper\,I).

The physical parameters of the bona-fide PMS stars that we have
identified are obtained as explained in the following. 

\subsection{Masses and ages}
\label{masag}

\begin{figure}
\centering
\resizebox{\hsize}{!}{\includegraphics[bb=40 180 490 600,width=16cm]
{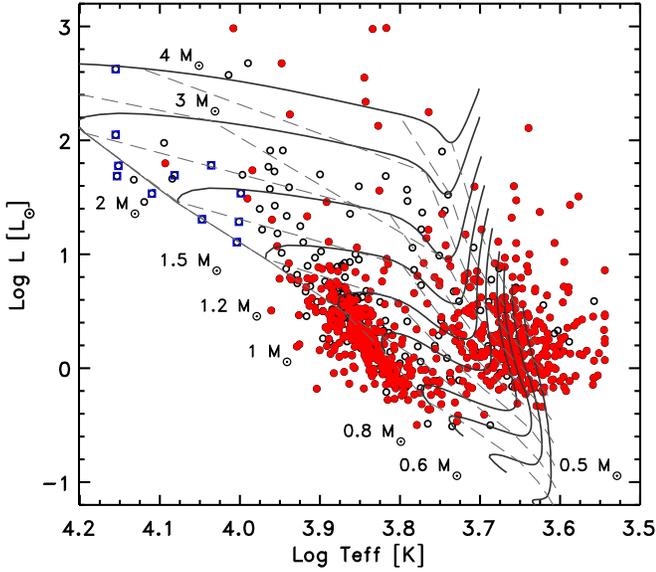}}
\caption{Hertzsprung--Russel diagram of the PMS candidate stars. All
objects shown here have excess emission in H$\alpha$ at the $4\,\sigma$
level or higher. Those indicated with filled circles (in red in the
online version) also have  $W_{\rm eq}(H\alpha) < -20$\,\AA\ or $<
-50$\,\AA\ for stars hotter than 10\,000\,K. Squares correspond to
objects with $W_{\rm eq}(H\alpha) > - 50$\,\AA\ and, as such, are
potential Be stars. Thick solid lines show the evolutionary tracks from
the Pisa group (Degl'Innocenti et al. 2008; Tognelli et al. 2011) for
metallicity $Z=0.002$ and masses from $0.5$ to 4\,\Msolar, as indicated.
The corresponding isochrones are shown as thin dashed lines, for ages of
$0.125$, $0.25$, $0.5$, 1, 2, 4, 8, 16 and 32\,Myr from right to left.
Note that the constant logarithmic age step has been selected in such a
way that the typical distance between isochrones is larger than the
photometric uncertainties.} 
\label{fig6}
\end{figure}

The radius $R_*$ comes from the luminosity and effective temperature of
the stars, which in turn follow from the observed colour $(V-I)$ and
magnitude ($V$), properly corrected for interstellar extinction as
explained in Section\,\ref{redde}. The stellar mass $M_*$ and age were
derived by comparing the location of each star in the
Hertzsprung--Russell (H--R) diagram of Figure\,\ref{fig6} with the PMS
evolutionary tracks and corresponding isochrones, via interpolation over
a finer grid than the one shown in the figure. We followed the
interpolation procedure developed by Romaniello (1998), which does not
make assumptions on the properties of the population, such as the
functional form of the IMF. {On the basis of the measurement errors,
and taking into account its position in the H--R diagram, this procedure
provides the probability distribution for each individual star to have a
given value of the mass and age (the method is conceptually identical to
the one presented recently by Da Rio et al. 2010). }

As for the theoretical models, we adopted  those of the Pisa group
(Degl'Innocenti et al. 2008; Tognelli et al. 2011), for metallicity
$Z=0.002$ or about $Z_\odot/8$. These new PMS tracks were specifically
computed for the mass range from $0.45$\,\Msolar to 5\,\Msolar with an
updated version of the FRANEC evolutionary code  (see Chieffi \&
Straniero 1989 and Degl'Innocenti et al. 2008 for details and Cignoni et
al. 2009 for an application). 

{As mentioned in the Introduction, we opted for a lower value of
the metallicity, namely $Z=0.002$ in place of the more customarily
adopted $Z=0.004$, because the corresponding tracks provide a better
match to the observations, particularly where they reach the MS. As
previously mentioned, $Z=0.002$ is still compatible with the metallicity
range found in the literature for the SMC and is in any case better 
suited to NGC\,346 than the value adopted in previous studies of its
PMS population that made use of the evolutionary tracks of Siess et al.
(2000) for $Z=0.01$. }

Figure\,\ref{fig6} reveals that the masses of the  bona-fine PMS stars
span over a decade, from $< 0.45$\,\Msolar for the coolest objects to
$\sim 4$\,\Msolar for the hottest ones. As mentioned above, some of the
objects with $T_{\rm eff} > 10\,000$\,K and $W_{\rm eq}(H\alpha) > 
-50$\,\AA\, could be Be stars and we have marked all objects of this
type with squares to distinguish them from the rest. Figure\,\ref{fig6}
also provides a more quantitative characterisation of the age difference
between the two populations of PMS stars that we identified earlier in
the CMD (Figure\,\ref{fig5}). The isochrones in the figure,
corresponding from right to left to ages of $0.125$, $0.25$, $0.5$, 1,
2, 4, 8, 16 and 32\,Myr, show that the young population has a median age
of $\sim 1$\,Myr, whereas the older bona-fide PMS stars span a wider age
range from $\sim 10$\,Myr to 30\,Myr, with a broad peak around 20\,Myr. 

\begin{figure}
\centering
\resizebox{\hsize}{!}{\includegraphics[bb=100 180 558 600]{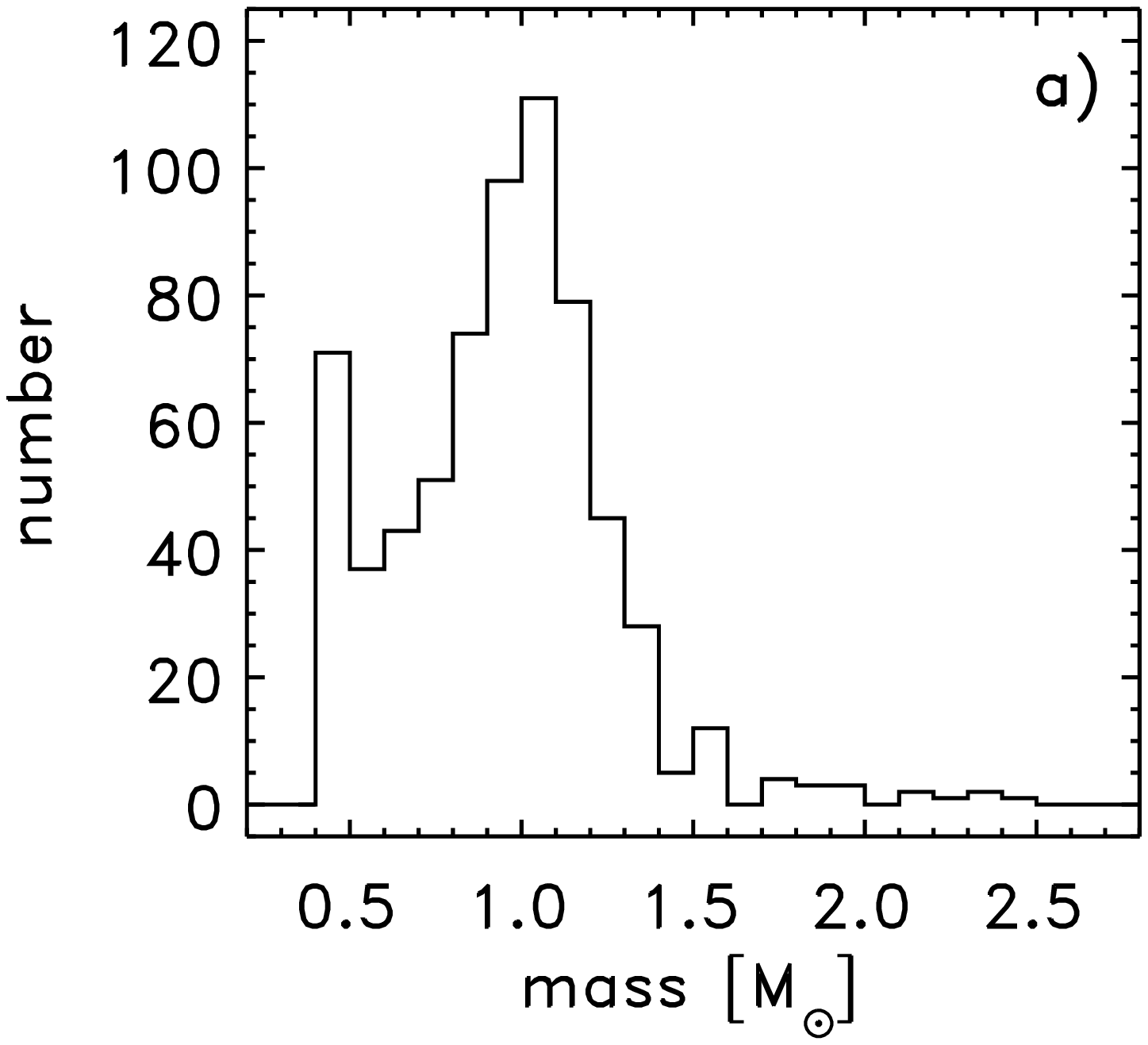}
\includegraphics[bb=100 180 558 600]{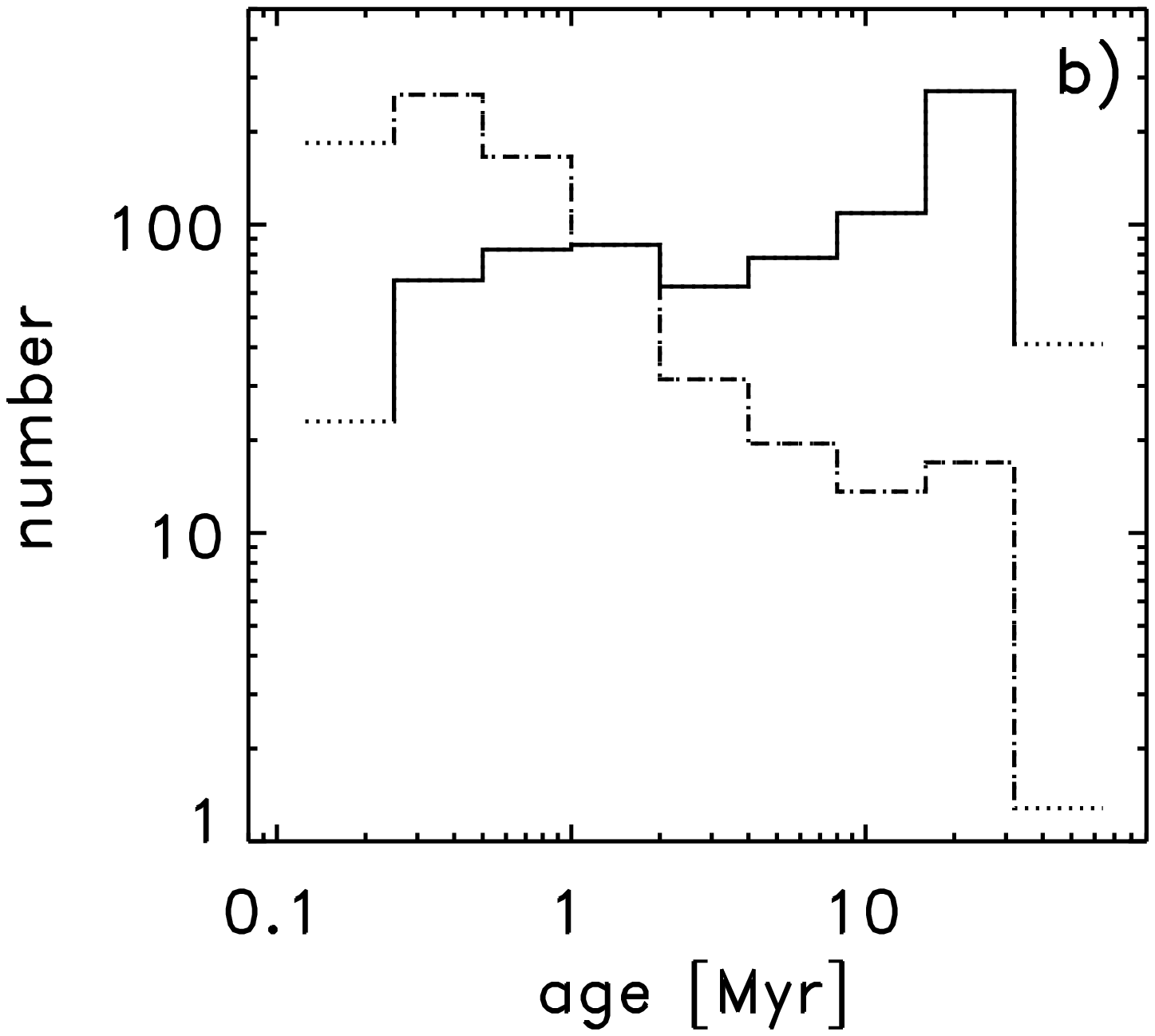}}
\resizebox{\hsize}{!}{\includegraphics[bb=100 180 558 600]{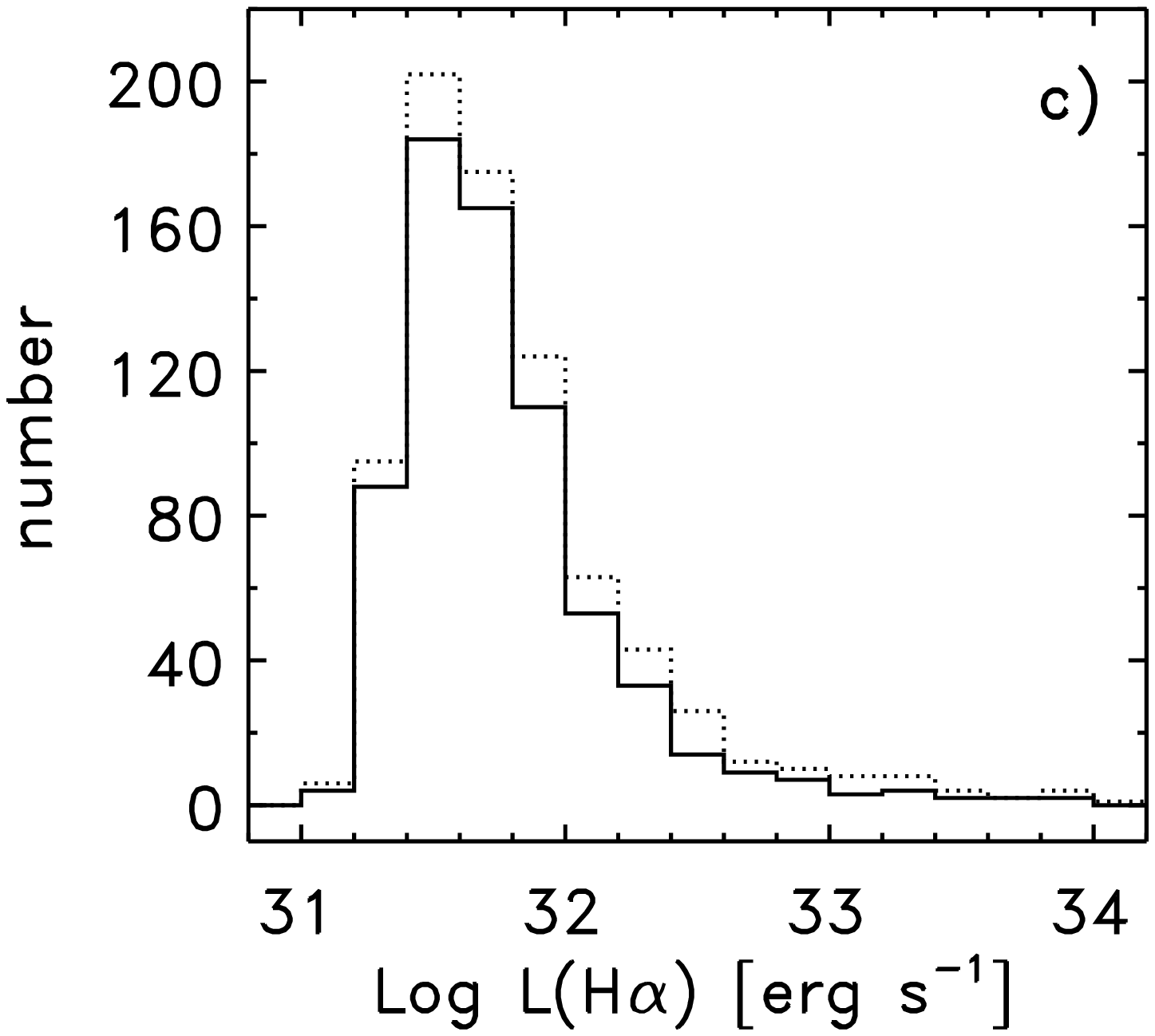}
\includegraphics[bb=100 180 558 600]{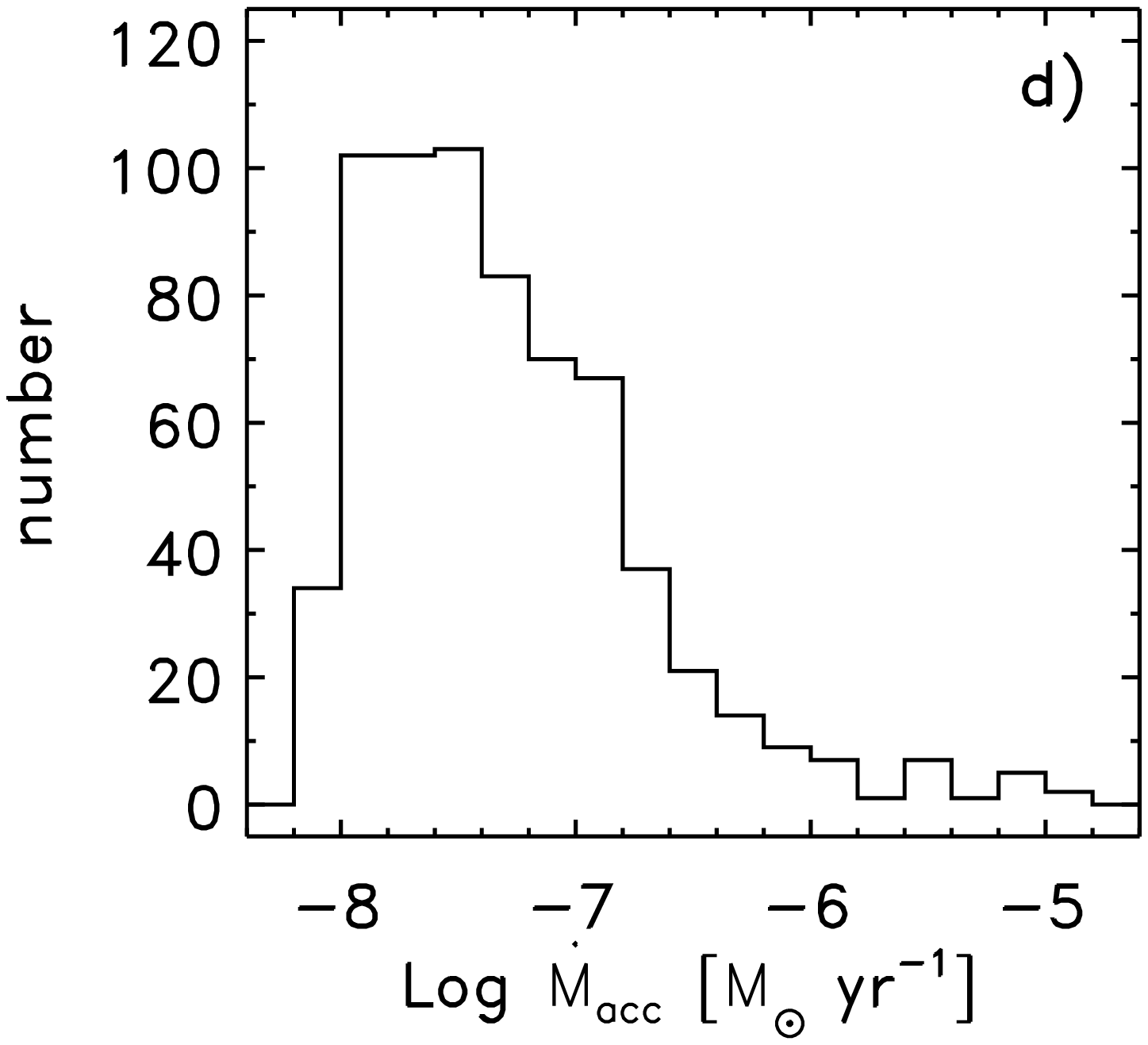}}
\caption{Histograms of the physical parameters of the
bona-fide PMS stars in the field of NGC\,346, namely mass ({\em a}), age
({\em b}), H$\alpha$ luminosity ({\em c}) and mass accretion rate ({\em
d}). The solid line refers to the 680 objects that also have an
equivalent width $W_{\rm eq}(H\alpha) < -20$\,\AA\ (or $< -50$\,\AA\ for
stars with $T_{\rm eff} > 10\,000$\,K) and a well defined mass. For
comparison, the dotted line in panel ({\em c}) refers to the  791 stars
with an H$\alpha$ excess at the $4\,\sigma$ level but without any
further restrictions on mass, age or equivalent width. } 
\label{fig7}
\end{figure}

We derived a reliable measure of the mass and age for all stars
indicated as thick dots in Figure\,\ref{fig6}, except for the coolest
objects that would have required a more uncertain extrapolation beyond
the limits of the theoretical models. Histograms with the mass and age
distribution for the 680 stars measured in this way are shown in
Figures\,\ref{fig7}a and \ref{fig7}b, respectively. As
Figure\,\ref{fig6} already implied, the  histogram in
Figure\,\ref{fig7}b shows that there is a remarkable paucity of stars
with ages around 4--8\,Myr. { In particular, the solid line in
Figure\,\ref{fig7}b provides a histogram of the age distribution with
constant logarithmic step (a factor of 2) and shows two peaks, at  $\sim
1$ and $20 - 30$\,Myr. The two peaks remain visible also when the number
of stars in each bin is rescaled by the size of the bin, as indicated by
the dot-dashed line (note that at the extremes of the distribution it
becomes more difficult to assign an age to the stars, so the first and
last bin are drawn with a dotted line to indicate a larger uncertainty).
In a companion paper (De Marchi,  Paresce \& Sabbi 2011), we discuss
the  age and spatial distributions of these stars in more detail and
show that between 30 and 10\,Myr ago a large number of PMS stars formed,
similar to that produced by the current burst, which however appears to
be much more intense.}

\subsection{Mass accretion rates}

Besides the H$\alpha$ luminosity $L(H\alpha)$ already discussed in the
previous section, the other physical parameters of interest in this work
are the accretion luminosity $L_{\rm acc}$ and the mass accretion rate
$\dot M_{\rm acc}$, that we derived following the procedure described in
Paper\,I. From an analysis of the literature values of $L_{\rm acc}$ and
$L(H\alpha)$ measurements as recently summarised by Dahm (2008) for PMS
stars in the Taurus--Auriga association, that paper concluded that the
ratio $L_{\rm acc} / L(H\alpha)$ can be assumed to be constant. The 
proportionality constant, obtained from an elementary fit
to the data in the compilation of Dahm (2008), is:

\begin{equation}
\log L_{\rm acc} = (1.72 \pm 0.47) + \log L(H\alpha)
\label{eq4}
\end{equation}
  
\noindent 
where the large uncertainty on the proportionality factor arises because
the two quantities were determined from non-simultaneous observations.
Considering that the intensity of the H$\alpha$ line from PMS sequence
stars is known to vary by about 20\,\% on a time  scale of a few hours
and by as much as a factor of 2 -- 3 in a few days (e.g.  Fernandez et
al. 1995; Smith et al. 1999; Alencar et al. 2001), the large spread can
be understood. However, as we will show in Section\,\ref{evolu}, the small
dispersion of the mass accretion rate with stellar age suggests an
uncertainty smaller than $0.25$.

The mass accretion rate is related to $L_{\rm acc}$ via the free-fall
equation, that links the luminosity released in the impact of the
accretion flow with the rate of mass accretion $\dot M_{\rm acc}$,
according to the relationship:

\begin{equation}
L_{\rm acc} \simeq \frac{G\,M_*\,\dot M_{\rm acc}}{R_*} \left(1 - 
\frac{R_*}{R_{\rm in}}\right)
\label{eq5}
\end{equation}

\noindent
where $G$ is the gravitational constant, $M_*$ and $R_*$ the mass and
photospheric radius of the star as derived above and $R_{\rm in}$ the
inner radius of the accretion disc. The value of $R_{\rm in}$ is rather
uncertain and depends on how exactly the accretion disc is coupled with
the magnetic field of the star. Following Gullbring et al. (1998), we
adopt $R_{\rm in} = 5\,R_*$ for all PMS objects. By substituting
Equation\,\ref{eq4} in Equation\,\ref{eq5}, the mass accretion rate
$\dot M_{\rm acc}$, in units of \Msolar~yr$^{-1}$, can be directly
expressed as a function of $L(H\alpha)$: 

\begin{eqnarray}
\log \frac{\dot M_{\rm acc}}{M_\odot {\rm yr}^{-1}} & = &-7.39 + \log
\frac{L_{\rm acc}}{L_\odot} + \log\frac{R_*}{R_\odot} -
\log\frac{M_*}{M_\odot} \\
& = & (-5.67 \pm 0.47) + \log\frac{L(H\alpha)}{L_\odot} + \log
\frac{R_*}{R_\odot} - \log \frac{M_*}{M_\odot} \nonumber
\label{eq6}
\end{eqnarray}

The histograms showing the distribution of $L(H\alpha)$ and $\dot M_{\rm
acc}$ as derived with Equation\,\ref{eq6} for the 680 bona-fide PMS
stars with well defined masses are shown in Figures\,\ref{fig7}c and
\ref{fig7}d, respectively. The median values of $L(H\alpha)$ and $\dot
M_{\rm acc}$ are, respectively,  $4.6 \times 10^{31}$\,erg/s (or
$0.012$\,L$_\odot$) and $3.9 \times 10^{-8}$\,\Msolar~yr$^{-1}$. The
latter value is about 50\,\% higher than the median mass accretion rate
of a population of 133 PMS stars that we detected in the field of
SN\,1987A (see Paper\,I) and this is fully consistent with the much
younger average age of the PMS stars in NGC\,346. 

Paper\,I provides an extensive discussion of the statistical and
systematic uncertainties involved in determining $\dot M_{\rm acc}$ with
this method and we refer the reader to that work for more details. The
typical combined statistical uncertainty on $\dot M_{\rm acc}$ for the
NGC\,346 field is 13\,\%. It is largely dominated by the uncertainty on
$L(H\alpha)$ of 10\,\% which in turn is mostly due to random errors in
the photometry (we note that this uncertainty is lower than for the
study of the field around SN\,1987A in Paper\,I because of the higher
quality of the ACS photometry). As for systematic effects, those on
$R_*$, $M_*$ and reddening add up to 15\,\%, but it is the rather
uncertain relationship linking $L(H\alpha)$ to the accretion luminosity
$L_{\rm acc}$ that plays a dominant role. We stress here that our method
adopts the calibration provided by Dahm (2008), which is based on a
spectroscopic study of Galactic PMS stars. If the uncertainty on the
relationship between $L(H\alpha)$ and $L_{\rm acc}$ is as large as
Dahm's (2008) work suggests (but see Section\,\ref{evolu}), it could possibly
cause a  systematic effect of up to a factor of 3 on the value of $\dot
M_{\rm acc}$ that we obtain.

This is the unfortunate consequence of our incomplete understanding of
the physics of the accretion process. Improving this situation would
require direct measurements of the bolometric accretion luminosity
$L_{\rm acc}$ and simultaneous measurements of the hot continuum excess
emission and of the emission line fluxes in H$\alpha$ or other lines
(e.g. Pa$\beta$ and Br$\gamma$). It is also true, however, that these
systematic effects plague in the same way all determinations of $\dot
M_{\rm acc}$ that make use of emission lines diagnostics. This includes
detailed studies of the profile and intensity of the H$\alpha$ line
(e.g. Muzerolle et al. 1998), since they rely on the same measurements
for their calibration. Therefore, it is possible and meaningful to
compare the values of $\dot M_{\rm acc}$ that we obtain here to one
another and to those derived with other methods, as we do in the
following section.

\begin{figure}
\centering
\resizebox{\hsize}{!}{\includegraphics{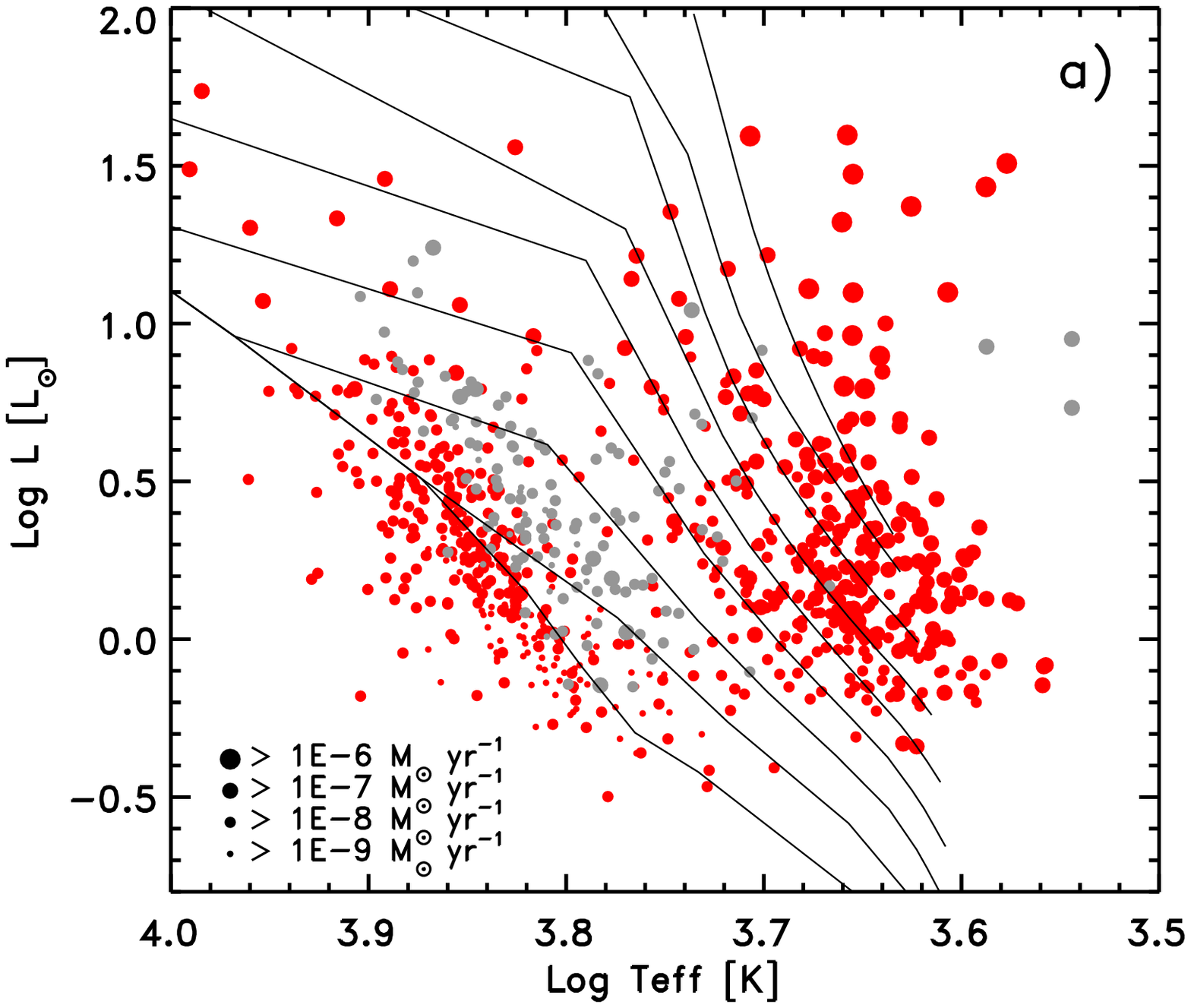}}
\resizebox{\hsize}{!}{\includegraphics{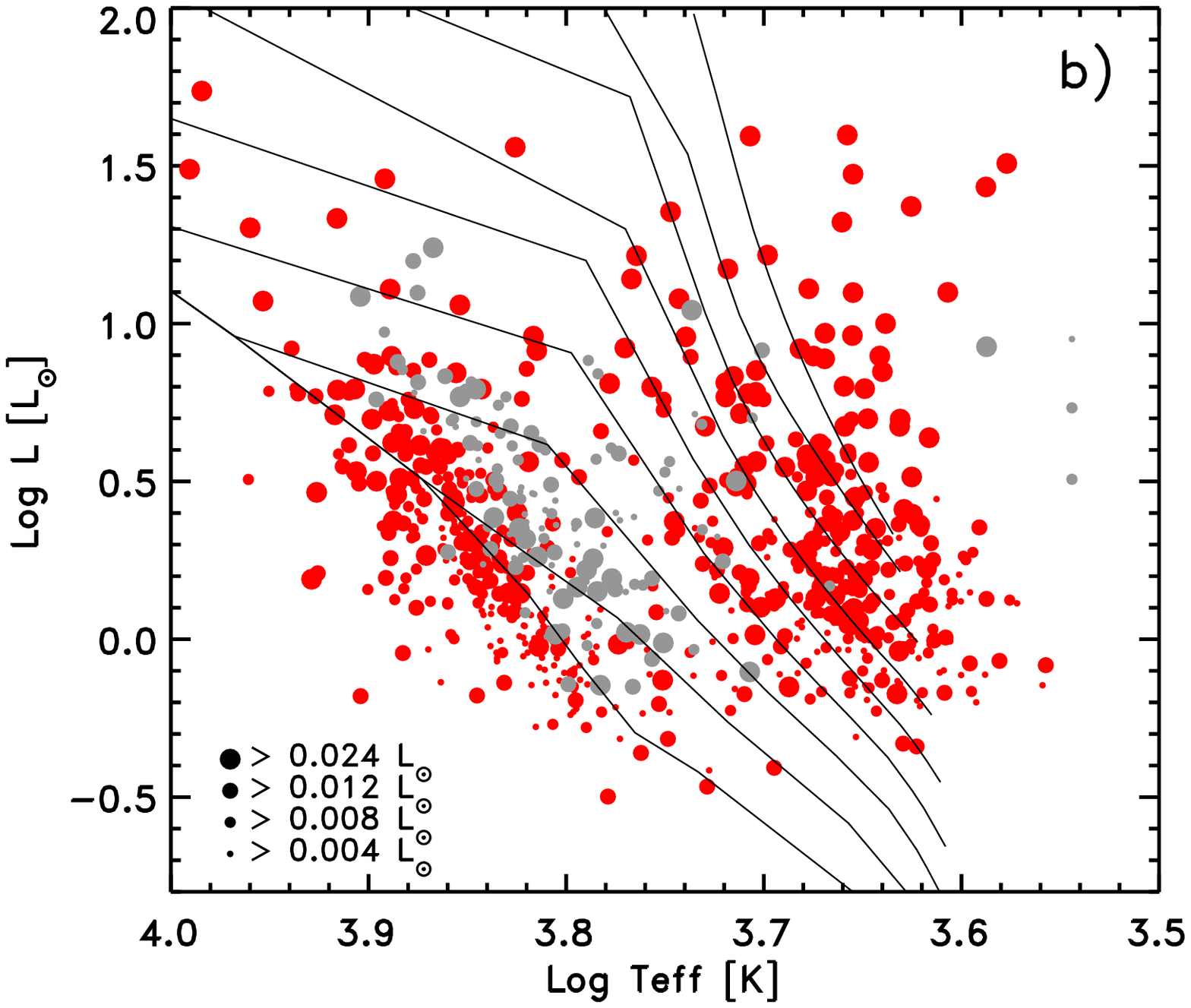}}
\resizebox{\hsize}{!}{\includegraphics{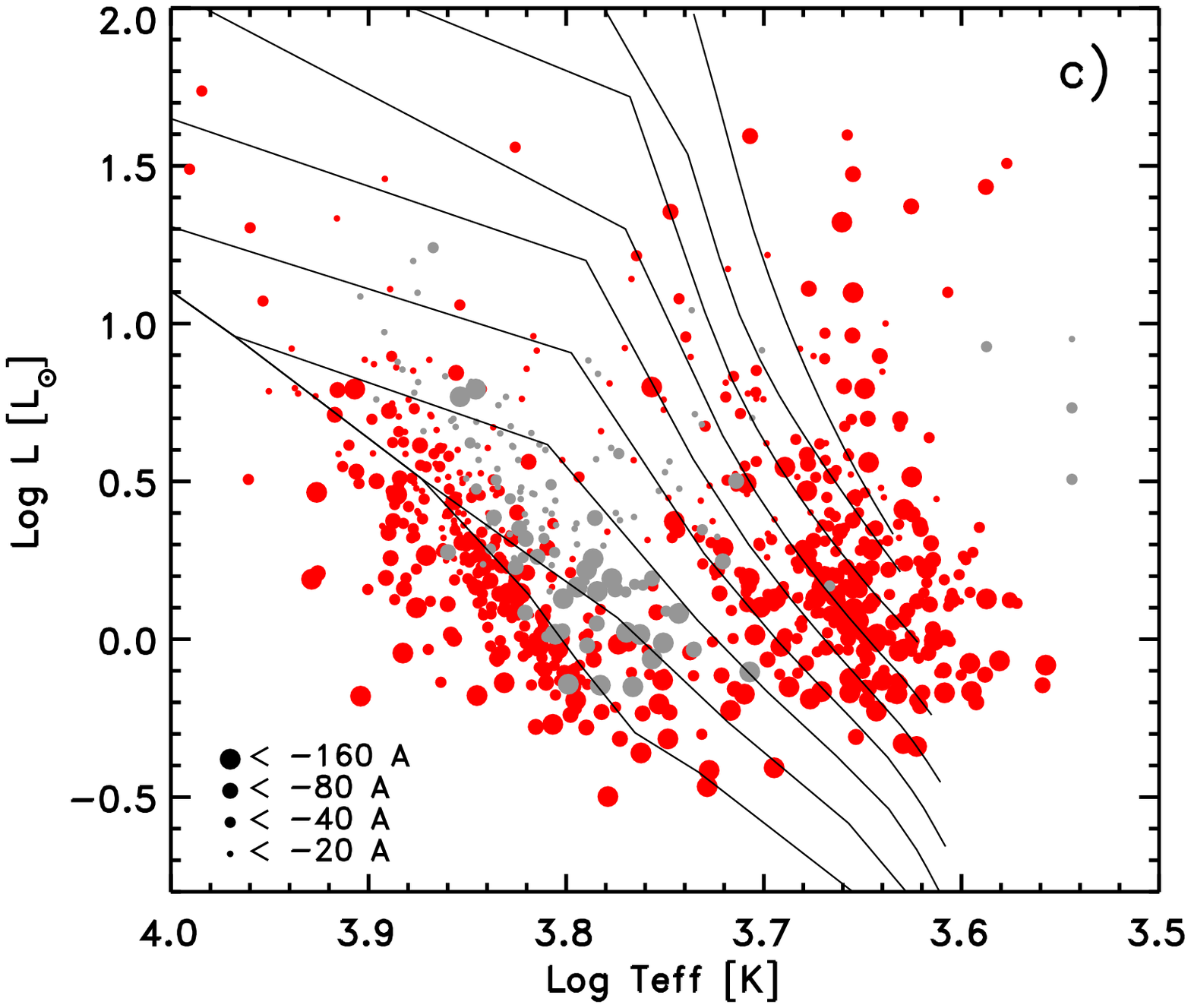}}
\caption{H--R diagrams of all 680 bona-fide PMS stars. In panels ({\em
a}), ({\em b}) and ({\em c}) the size of the symbols is set to be
proportional respectively to $\dot M_{\rm acc}$, $L(H\alpha)$ and 
$W_{\rm eq}(H\alpha)$, according to the legends. The darker symbols (in
red in the online version) correspond to NGC\,346 (same as
Figure\,\ref{fig6}), while the lighter ones (gray in the online version)
represent the bona-fide PMS stars around SN\,1987A. The isochrones
corresponds from right to left to ages from $0.125$\,Myr to 32\,Myr with
a constant logarithmic step and are for metallicity $Z=0.002$.
Therefore, isochrones are only applicable to the NGC\,346 data.} 
\label{fig8}
\end{figure}

\section{Evolution of the mass accretion rate}
\label{evolu}

In this section we investigate how $\dot M_{\rm acc}$ varies with
stellar parameters such as stellar mass and age. It is important to
underline that this analysis is based only on stars that, at the time of
the observations, were undergoing active mass accretion as revealed by
H$\alpha$ excess emission above our detection threshold. Actually, PMS
objects are known to display large variations in their H$\alpha$
emission over hours or days (e.g. Fernandez et al. 1995; Smith et al.
1999; Alencar et al. 2001), with only about one third of them at any
given time being active H$\alpha$ emitters with $W_{eq}(H)< -10$\,\AA\
in the field around SN\,1987A (Panagia et al. 2000). 

In fact, a similar requirement of a high accretion activity is common to
{\it all} studies in which PMS stars are identified based on
single-epoch observations, including those conducted spectroscopically
on T Tauri objects in the Milky Way (e.g. Sicilia--Aguilar et al. 2006).
This implies that the accretion rates measured for the detected objects
will, in fact, represent the highest values possible for that class of
objects, while many more stars may well be accreting at more modest
rates.  As a result, the true {\it average} accretion rate appropriate
for a given class of objects will be just a fraction of the values
measured for the members of that class that can be detected. On the
other hand, since all studies made with different techniques suffer from
the same kind of selection on the accretion efficiency, a direct
comparison between their results is still possible and meaningful. 

\subsection{A qualitative analysis}

Let's start with a qualitative analysis of the H--R diagram of the
bona-fide PMS stars shown in Figure\,\ref{fig8}a, where the size of the
symbols is set to be proportional to $\dot M_{\rm acc}$ according to the
legend. Besides the 680 bona-fide PMS stars in NGC\,346 (darker dots,
red in the online version) we have included in the figure for comparison
also the 133 bona-fide PMS stars in the field of SN\,1987A discussed in
Paper\,I (lighter dots, gray in the online version). The theoretical PMS
isochrones are the same as shown in Figure\,\ref{fig6} and correspond to
a metallicity $Z=0.002$ and to ages from $0.125$\,Myr to 32\,Myr
increasing from the right with a constant logarithmic step of a factor of
2. While not directly applicable to the PMS stars in the SN\,1987A
field, due to the different metallicity, these isochrones serve
nonetheless the purpose to illustrate the effects of increasing PMS age.

It is noteworthy that the mass accretion rate varies by over three
orders of magnitude across the H--R diagram in a very regular way:
younger stars, which are farther away from the MS, have typically higher
mass accretion rates, as expected from the theory of stellar evolution.
Also the stellar mass appears to affect the value of $\dot M_{\rm acc}$,
in the sense that less massive stars tend to have smaller accretion
rates. It is also quite remarkable that even PMS stars very close to the
location of the zero-age MS (ZAMS), which for the masses of interest
here can be considered equivalent to the 32\,Myr PMS isochrone, are
still undergoing substantial mass accretion. 

To verify that this is not a spurious effect due to errors in our
derivation of $\dot M_{\rm acc}$, we show in Figure\,\ref{fig8}b the
same H--R diagram but with the size of the symbols proportional to the
measured H$\alpha$ luminosity. There one sees that $L(H\alpha)$ can be
rather large for objects at or near the ZAMS, suggesting that the
accretion process remains relatively active when the star approaches the
MS. More precisely, this demonstrates that, even if $\dot M_{\rm acc}$
decreases as a PMS star approaches the MS, its shrinking radius allows
the star to extract more potential energy per unit rate of mass
accretion (i.e. for a given rate in g\,cm$^{-1}$), thereby making the
object easier to detect in H$\alpha$. Finally, Figure\,\ref{fig8}c shows
that the $EW(H\alpha)$ values of PMS objects near or at the MS are 
typically much larger than those caused by chromospheric activity for
stars in this mass range (see e.g. Ulmschneider 1979; Pasquini et al.
2000), confirming that they are bona-fide young objects.

\subsection{A quantitative analysis}

\begin{figure}
\centering
\resizebox{\hsize}{!}{\includegraphics[bb=95 180 550 600]{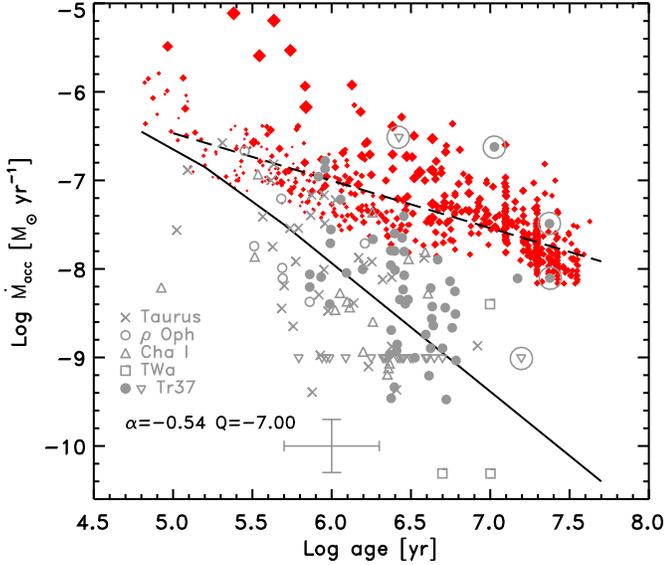}}
\caption{Mass accretion rate as a function of stellar age. The 680
bona-fide PMS stars that we identified in NGC\,346 are indicated as
diamonds, with a symbol size proportional to the stellar mass. The
dashed line is the best fit to their distribution, with slope $\alpha
=-0.54$ and intercept $Q=-7$ (at 1\,Myr). The other symbols (see legend)
correspond to the Galactic T-Tauri stars studied by Sicilia--Aguilar et
al. (2006), whose distribution shows a large scatter but is nonetheless
consistent with the models of viscous disc evolution  of Hartmann et al.
(1998) represented by the solid line.}
\label{fig9}
\end{figure}

For a more quantitative analysis of the evolution of the mass accretion
rate, we turn to Figure\,\ref{fig9} where the variation of $\dot M_{\rm
acc}$ is shown as a function of stellar age. All 680 bona-fide PMS
stars in NGC\,346 are indicated as diamonds, with a symbol size
proportional to their mass. Besides being very homogeneous, this sample
is about five times larger than all $\dot M_{\rm acc}$ measurements
today available for Galactic PMS objects taken together (e.g.
Sicilia--Aguilar et al. 2006). Furthermore, it spans over a decade in
mass and two orders of magnitudes in age while including objects from
the same star forming region with very accurate and uniform
measurements. All this makes the sample particularly well suited for a
quantitative analysis of the dependence of the mass accretion rate on
stellar parameters.

The dashed line in Figure\,\ref{fig9} represents the best fit to the
evolution of the mass accretion rate in NGC\,346, with a slope $\alpha
=-0.54$ and intercept $Q=-7$ (at 1 \,Myr). The fit appears considerably
less steep than the prediction of Hartmann et al. (1998; see also
Muzerolle et al. 2000), based on models of viscous disc evolution and
represented here by the solid line, which however seems  to adequately
reproduce the trend of decreasing $\dot M_{\rm acc}$ with stellar age
for low-mass Galactic T-Tauri stars as compiled by Sicilia--Aguilar et
al. (2006) and also shown in the figure. The different slopes of the
observed and model relationships translate into predicted $\dot M_{\rm
acc}$ values that are about two order of magnitudes smaller than what we
actually observe for an age of 10\,Myr. 

This discrepancy had already been noticed in the field of SN\,1987A
studied in Paper\,I, although it is pointed out there that the $\dot
M_{\rm acc}$ value measured at 10\,Myr is in excellent agreement with
that of the  most massive PMS stars (with spectral type G) in the
Trumpler\,37 sample of Sicilia--Aguilar et al. (2006) shown as large
circles in the figure. Even though the age of the G-type stars in
Trumpler\,37 is very uncertain (Sicilia--Aguilar et al. 2006) and could
be much younger (J. Muzerolle,  priv. comm.; see also Hartmann 2003), we
cannot exclude that the slower decline of $\dot M_{\rm acc}$ with time
that we observe in NGC\,346 and around SN\,1987A could be due to the
typically higher mass of our objects compared with that of the Galactic
T Tauri stars that were the subject of recent studies. A dependence of
$\dot M_{\rm acc}$ on the stellar mass $M$ is indeed expected and could
be as steep as $\dot M_{\rm acc} \propto M^2$ according to Muzerolle et
al. (2003; 2005), Natta et al. (2004; 2006) and Calvet et al. (2004),
although the uncertainties are large (see also Clarke \& Pringle 2006
for a different interpretation).

\begin{figure}
\centering
\includegraphics[scale=0.47,bb=30 160 530 660,clip=true]{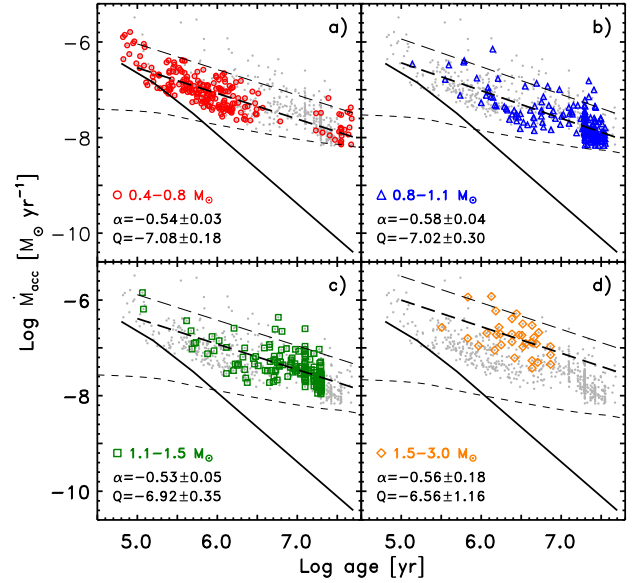}
\caption{Same as Figure\,\ref{fig9} but for stars in four mass groups,
as indicated in each panel. We also provide the slope $\alpha$ and 
intercept $Q$ (at 1\,Myr) of the best fits (thick long-dashed lines).
The thin short-dashed lines correspond to our $L(H\alpha)$ detection
limits, which appear to not significantly affect the determination of the
slopes, except possibly for the least massive stars at ages $>
10$\,Myr.} 
\label{fig10}
\end{figure}

Owing to the large size and wide mass range of the objects in our
sample, we can investigate the mass dependence of $\dot M_{\rm acc}$ in
a very robust way. As a first step, we divide out sample in four mass
groups, namely $0.4 - 0.8$\,\Msolar, $0.8 - 1.1$\,\Msolar, $1.1 -
1.5$\,\Msolar and $1.5 - 3$\,\Msolar, and plot for each one separately 
the temporal evolution of $\dot M_{\rm acc}$ in the four panels of 
Figure\,\ref{fig10}. The slope $\alpha$ and intercept $Q$ (at 1\,Myr) of
the best linear fit to the data, according to the relationship $\log
\dot M_{\rm acc} = \alpha \times \log({\rm age}) + Q$, are given in each
panel together with their uncertainty. It appears that the slopes of
the four mass groups are in very good agreement with one another, within
the uncertainty of the fits, but remain significantly different from the
trend predicted by Hartmann et al. (1998) and Muzerolle et al. (2000)
for viscous disc evolution (solid lines). 

In principle, the shallower slopes of our fits could be affected by our
very strict criteria for selecting bona-fide PMS stars. Requiring that
the H$\alpha$  excess be at the $4\,\sigma$ level or higher and that
$W_{\rm eq} < -20$\,\AA\ will set a lower limit to the H$\alpha$
luminosity that we can accept as significant and, therefore, to the
smallest $\dot M_{\rm acc}$ value that we can detect at a given mass or
age. If this lower limit is too high, we could be ignoring too many
objects with low mass accretion rate, particularly at older ages, and
therefore skew the measurement of the slope towards low values. On the
other hand, it appears that our detection limits (shown by the thin
short-dashed lines for the average mass of each group) do not
significantly affect the determination of the slope, except possibly for
the least massive stars at ages $> 10$\,Myr. For this group, however,
the slope is determined by the much more numerous young stars. 

More importantly, the upper envelopes of the distribution of all four
mass groups appear to be fully consistent with the slope of the best
fit. This is shown graphically by the thin long-dashed lines that
represent the best-fitting line shifted vertically by $0.5$\,dex. While
our conservative limits on the H$\alpha$ excess emission may prevent us
from detecting objects with low $L(H\alpha)$, no selection effects are
to be expected for stars with high H$\alpha$ luminosity. Since the
distribution of these objects is consistent with the slope of the best 
fit, the latter must be correct. Therefore, within the uncertainties
quoted in Figure\,\ref{fig10}, we can conclude that all four mass groups
are compatible with the same slope $\alpha \simeq -0.6$ (formally
$\alpha=-0.55 \pm 0.02$). 


An interesting feature of Figure\,\ref{fig10} is that the spread of the
$\log \dot M_{\rm acc}$ data values around the best fitting lines is
small, typically $\sim 0.25$\,dex (rms). This proves quite convincingly
that the true uncertainty in the relationship between $\log L_{\rm acc}$
and $\log L(H\alpha)$ given by Equation\,\ref{eq4} cannot be as large as
the $0.47$\,dex value suggested by our elementary fit to the data in the
compilation of Dahm (2008), but it has to be intrinsically smaller than
the $\sim 0.25$\,dex factor that we observe in our own NGC\,346 data. We
conservatively assume $0.25$\,dex (or a factor of $\la 1.8$ in linear
units).


Having established that the data in all four panels of
Figure\,\ref{fig10} are consistent with the same slope $\alpha \simeq
-0.6$, we can use the intercept $Q$ at 1\,Myr to study how $\dot M_{\rm
acc}$ varies with the stellar mass $m$. It appears that the value of $Q$
in the four panels of Figure\,\ref{fig10} decreases with increasing mass
approximately as $m^{-1}$. The formal uncertainty on $Q$ may seem too
large to reach this conclusion, but the scatter is due to the grouping
of stars of different mass in the same bin. It is, therefore, more
appropriate to perform a multivariate fit to the three variables
simultaneously ($\dot M_{\rm acc}$, age and mass). We opt for a linear
multivariate least-square fit of the type:

\begin{equation}
\log \dot M_{\rm acc} = a \times \log t + b \times \log m + c
\label{eq7}
\end{equation}

\noindent
where $t$ is in years, $m$ the mass in solar units and $c$ is a 
constant. The formal best fit gives $a=-0.61 \pm 0.02$, $b=0.90 \pm
0.10$ and $c=-3.27 \pm 0.24$. However, since the set of parameters 
$a=-0.6$, $b=1$ and $c=-3.3$ would give almost equally small residuals,
we will take them hereafter as a good as well as convenient
representation of the dependence of $\dot M_{\rm acc}$ on stellar mass
and age in NGC\,346. 

\begin{figure}
\centering
\includegraphics[scale=0.55,bb=100 180 558 600]{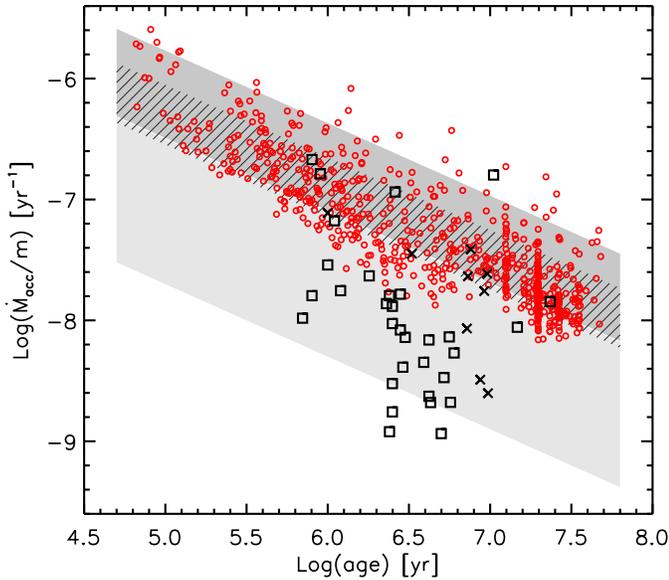}
\caption{Run of $\dot M_{\rm acc}/m$ as a function of age for the PMS
objects in NGC\,346 (circles) and for Galactic stars in Taurus (crosses)
and in Trumpler 37 (squares). The hatched band show the best fit ($\pm
1\,\sigma$) to the NGC\,346 data, the dark-shaded band is the fit to the
PMS stars in the 30\,Dor region as derived in Paper\,III and the
light-shaded band is the fit to the Galactic stars (including also the
Orion Nebula Cluster; Da Rio et al. in preparation).} 
\label{fig11}
\end{figure}

\subsection{Comparison with other star forming regions}

From this analysis it appears that the NGC\,346 data are not consistent
with a steep dependence of $\dot M_{\rm acc}$ on the stellar mass of the
type reported by Muzerolle et al. (2003) and Calvet et al. (2004) in
nearby star forming regions. However, the gentler decline of $\dot
M_{\rm acc}$ with mass that we find here is fully compatible with that
obtained in a study of about 900 PMS stars recently identified in the 
regions around 30\,Dor (Spezzi et al., in preparation, Paper III) as
well as with the preliminary results of a ground-based study of about
500 PMS objects in the Orion Nebula Cluster (Da Rio et al., in
preparation). In all these cases, values of $a=-0.6$ and $b=1$ in
Equation\,\ref{eq7} give a good fit to the data. 

This suggests that we could use the same functional form and parameter
values to fit the observed evolution of $\dot M_{\rm acc}$ for Galactic
PMS stars and compare the result to the data in the Magellanic Clouds.
We do so in Figure\,\ref{fig11}, where we show the run of the {\it
specific} mass accretion rate, defined as $\dot m_{\rm}=\dot M_{\rm
acc}/m$, as a function of time for the PMS objects in NGC\,346 (circles)
and for Galactic stars in Taurus (crosses; from Calvet et al. 2000) and
Trumpler 37 (squares; from Sicilia-Aguilar et al. 2006). The hatched
band represents the best fit to the NGC\,346 data, the dark-shaded band
is the fit to the PMS stars in the 30\,Dor region as derived in
Paper\,III, whereas the light-shaded band refers to Galactic stars
(including also the Orion Nebula Cluster; Panagia et al. in
preparation). The hatched and solid bands correspond to the $\pm 1 \,
\sigma$ scatter (i.e. the range from 17 to 83 percentile) and indicate
that the SMC and LMC data are compatible with one another, but that the
difference with Galactic PMS stars is significant. 

The logarithmic value of the specific mass accretion rate at 1\,Myr, 
$q=\log \dot m(1\,{\rm Myr})$ calculated using the best fitting line as
given in Equation\,\ref{eq7}, can serve as a measure of this difference.
$q$ takes on the values of $-6.9 \pm 0.2$, $-6.7 \pm 0.4$ and $-7.7 \pm
0.6$ respectively for NGC\,346, the 30\,Dor region and the Galaxy. As
already pointed out in Paper\,I, the higher $\dot M_{\rm acc}$ values of
PMS stars in the LMC compared to Galactic objects of the same age could
result from the lower metallicity of the clouds. The radiation pressure
of the forming star is expected to be less strong on lower-metallicity
disc material and this will delay the dissipation of the disc, thereby
keeping the accretion process active for a longer time. This scenario is
compatible with the PMS stars in the Galaxy having a smaller value of
the specific mass accretion rate $q$ than those in the Magellanic
Clouds. 

A possibility would be that the shallower photometric depth of the LMC
data may cause the $L(H\alpha)$ detection threshold for PMS stars to be
higher (see Paper\,III for more details) and thus could skew the median
value of $\dot M_{\rm acc}$. We will address this possibility in detail
through the analysis of recently obtained HST observations of the same
region (De Marchi et al., in preparation). Alternatively, it is also
possible that below a certain metallicity threshold the radiation
pressure on the disc material no longer changes. Investigating this
possibility requires the analysis of larger and homogeneous data set of
$\dot M_{\rm acc}$ measurements in low-metallicity environments. 

\subsection{Mass growth during the PMS phase}

Finally, we would like to highlight an interesting corollary result
stemming from the large mass accretion rate values that we find for the
objects in NGC\,346. Using the rates as expressed by Equation\,\ref{eq7}
we can estimate the total mass accreted by a star over the time span
from birth to the ZAMS. Already in 1993 Palla \& Stahler (1993) showed
that the timescale to reach the ZAMS is a strong function of the stellar
mass. Fitting the more modern evolutionary calculations (Degl'Innocenti
et al 2008; Tognelli et al 2011) for metallicities {\small
$1/2-1/8$}\,Z$_\odot$ and stellar masses in the range $0.6$ up to
4\,\Msolar, to within 8\,\% accuracy we can express that time in years
as: 

\begin{equation}
t_{\rm ZAMS} \simeq 20 \times 10^6 ~~ (Z/0.002)^{0.5} ~~ m^{-2.3}.
\label{eq8}
\end{equation}

Equation\,\ref{eq7} provides an analytical expression for the average
mass accretion rate of the detected stars. On the other hand, there must
be an appreciable fraction of PMS objects that, at any given time, are
not detectable as strong H$\alpha$ emitters. Therefore, in order to
properly estimate the amount of accreted mass over the PMS phase we have
to make allowance for the appropriate duty cycle. While doing this
would require a detailed knowledge of the time evolution of the
H$\alpha$ emission for a large sample of PMS stars in this region, for
the sake of simplicity we will assume that the evolution consists in a
series of two-level transitions that recur in time. In this simplified
model the high state value corresponds to the average value measured for
the detected PMS stars and the low state of the H$\alpha$ undetected
stars is zero. Thus, the true mass accretion rate of any star of mass
$m$ at the time $t$ will simply be the average value measured for the
detected PMS stars multiplied by an efficiency factor $\phi$ that in
principle is a function of the stellar mass, the time and the
metallicity of the region. 

An approximate value of this function is given by the ratio of the
number of PMS stars with detected H$\alpha$ emission and the total
number of candidate PMS stars as defined by their location in the H--R
diagram. In a companion paper, De Marchi, Panagia \& Sabbi (2011) show
that for NGC\,346 such a ratio is essentially constant over time up to
PMS ages of 8\,Myr at a level of $<\phi> \simeq 0.28$. Note that similar
values have been found in the field of SN\,1987A for PMS stars with
median age of 14\,Myr ($<\phi> \simeq 0.32$; Panagia et al 2000).
Therefore, adopting a constant value of $<\phi>=0.28$ for PMS stars in
NGC\,346  allows us to integrate the accretion rate over time from 0 up
to $t_{\rm ZAMS}$ in order to estimate the total amount of mass (in
\Msolar units) accreted by these objects during their PMS phase. From
Equations\,\ref{eq7} and 8 we can express it in units of \Msolar as
follows:

\begin{eqnarray}
M_{acc} &=& \int_0^{t_{ZAMS}} \hspace*{-0.5cm}{\rm <}\phi{\rm >} ~~ 
            \dot M_{\rm acc} ~ dt \\ 
        &=& \int_0^{t_{ZAMS}} \hspace*{-0.5cm}{\rm <}\phi{\rm >} ~~
            t^a ~ m^b ~ 10^c ~ dt \nonumber  \\ 
        &=& 4 \times 10^{-4} ~ {\rm <}\phi{\rm >} ~ m  
            \int_0^{t_{ZAMS}} \hspace*{-0.5cm} t^{-0.6} \,dt  \nonumber \\
        &=& 4 \times 10^{-4} ~ {\rm <}\phi{\rm >} ~ m ~ 
            \frac{t_{ZAMS}^{0.4}}{0.4} \nonumber \\
        &=& 0.83 ~ {\rm <}\phi{\rm >} ~ m^{0.08} \nonumber \\
        &=& 0.23 ~ m^{0.08} \nonumber  
\label{eq9}
\end{eqnarray}
\noindent
showing that the amount of accreted mass is virtually insensitive to the
stellar mass. In the range  $0.4$\,\Msolar$<m< 4$\,\Msolar as studied
here, Equation\,\ref{eq9} implies $M_{\rm acc} = 0.23 \pm 0.03$\,\Msolar.

Taken at face value, such a high $M_{\rm acc}$ figure would imply that a
considerable amount of mass is accreted during the PMS phase, {at
least in low-metallicity environments such as the Magellanic Clouds.} If
this is the case, the PMS evolution of moderate-mass stars, say $<
2$\,\Msolar, should be reconsidered and recalculated taking into account
the high $\dot M_{\rm acc}$ values. One of the effects of Equation\,9 is
that the appropriate $t_{\rm ZAMS}$ for a star that reaches the MS with
a given mass will be longer than the value estimated by evolutionary
models that assume $\dot M_{\rm acc} \equiv 0$ after the first few
$10^4$\,yr (e.g. Baraffe et al. 2009). Since the star had an appreciably
lower mass for a very long time (several Myr) at the early epochs, it
was necessarily evolving more slowly.

Extrapolating these findings to lower masses than those that we have reached
here (i.e. $< 0.4$\,\Msolar) would carry the implication that no star could
form with a mass below a suitable minimum value. For NGC\,346 this would
correspond to a lower cut-off of the initial mass function at $\sim
0.2$\,\Msolar. In the Milky Way, where $\dot M_{\rm acc}$ is about five times
lower but the metallicity is about 8 times higher so that the PMS evolutionary
times are a factor of about $2.8$ longer, the cut-off would occur at $\sim
0.06$\,\Msolar. 

On the other hand, it is also possible that for low-mass stars  
circumstellar discs are eroded in a finite time that is appreciably 
shorter than their $t_{\rm ZAMS}$. For example, if for stars below
1\,\Msolar we were to limit the integrals in Equation\,9  between 0
and the smaller of $t_{\rm ZAMS}$ and 20\,Myr, the accreted mass would
decrease proportionally to the mass, i.e.

\begin{eqnarray}
M_{\rm acc} \simeq 0.23 ~~ m & &
\hspace*{2cm} {\rm for} \hspace*{0.5cm} m < 1\,{\rm M}_\odot.
\label{eq11}
\end{eqnarray}

In practice this would still require some refinements to models of
stellar evolution in the PMS phase, but the increase in mass would never
be higher than 20\,\% of the final ZAMS mass. Note that the choice of
20\,Myr as the upper limit for the integration stems from the fact that
near the MS we still find 1\,\Msolar stars that have strong H$\alpha$
excess, thus indicating the presence of considerable accretion. Deeper
observations of NGC\,346 reaching well below 1\,\Msolar on the MS can
easily show whether the discs of lower-mass objects are indeed eroded
before they reach the ZAMS. It would be sufficient to verify whether
strong H$\alpha$ excess emission is still shown by stars smaller than
1\,\Msolar near the MS, or, alternatively, whether for objects of these
lower masses strong H$\alpha$ emission is only seen at younger ages,
when they are still well above the MS. If at these masses the discs are
indeed eroded before reaching the ZAMS, H$\alpha$ emission should
systematically taper off for objects older than a cut-off age. In any
case, these new results can have important implications for some of the 
assumptions and boundary conditions used in theories of star formation
and early evolution.

\section{Summary and conclusions}
\label{summa}

We have studied the properties of the stellar populations in the field
of the NGC\,346 cluster in the SMC as observed with the ACS camera on board
the Hubble Space Telescope (Sabbi et al. 2007). To this aim, we have
employed a novel, self-consistent method that allows us to reliably
identify PMS stars undergoing active mass accretion, regardless of their
age. The method, fully described in Paper\,I, does not require
spectroscopy and combines broad-band $V$ and $I$ photometry with
narrow-band $H\alpha$ imaging to identify all stars with excess
H$\alpha$ emission and allows us to derive the accretion luminosity
$L_{\rm acc}$ and mass accretion rate $\dot M_{\rm acc}$ for all of
them. The main results of this work can be summarised as follows.

\begin{enumerate}

\item
From the photometry of Sabbi et al. (2007), we have selected all objects
in a region of $200\arcsec \times 200\arcsec$ around the centre of
NGC\,346 that have a combined mean error $\delta_3$ in the three bands
($V$, $I$ and H$\alpha$) of less than $0.1$\,mag, as defined in
Equation\,\ref{eq1}. A total of 18\,764 stars satisfy this condition.

\item 
The CMD reveals the presence of both young and old populations, but the
conspicuous broadening of the upper MS proves that there is also
variable extinction. We have determined the reddening towards each of
these stars from their colour and magnitude displacements with respect
to a 4\,Myr isochrone for metallicity $Z=0.002$ and a distance modulus
of $18.92$. We found that extinction is in the range $0.1 < E(V-I) <
0.25$ or $0.27 < A_V < 0.67$  (respectively 17\,\% and 83\,\% limits),
with a median value of $E(V-I)=0.16$ or $A_V=0.43$. We have then used
the reddening values towards each of the selected stars to derive a
reddening correction for all objects in their vicinity. 

\item 
Following the method developed in Paper\,I, we have identified 791 PMS
candidates as being objects with H$\alpha$ excess above the $4\,\sigma$
level with respect to the reference provided by normal stars. Their
average H$\alpha$ luminosity is $2.7 \times 10^{31}$\,erg\,s$^{-1}$ or
$\sim 10^{-2}$\,L$_\odot$. We have also determined the equivalent width
of the H$\alpha$ emission line of these objects and have classified as
bona-fide PMS stars all those with $W_{\rm eq} < -20$\,\AA\ (or 
$<-50$\,\AA\ for stars with $T_{\rm eff} > 10\,000$\,K). These conditions
guarantee that our sample is free from contamination due to the
chromospheric activity of older objects or to the rotational winds of Be
stars. A total of 694 objects satisfy these conditions.

\item
By comparing the locations of these objects in the H--R diagram with the
PMS evolutionary models of the Pisa group for metallicity $Z=0.002$, we
have been able to derive an accurate value of the mass and age of 680
bona-fide PMS stars (note that previous determinations of these
parameters for candidate PMS stars in NGC\,346 that made use of
evolutionary models for $Z=0.01$ are necessarily less accurate). The
masses of these objects range from $0.4$\,\Msolar to $4$\,\Msolar, with
an average value of $\sim 1$\,\Msolar. Their ages show a clear bimodal
distribution with two peaks at $\sim 1$\,Myr and $\sim 20$\,Myr and very 
few objects around $\sim 8$\,Myr, revealing the presence of two distinct 
but equally populous generations of stars. We address in a companion 
paper (De Marchi, Panagia \& Sabbi 2011) the properties of and the 
relationships between the two generations.   

\item 
From the H$\alpha$ luminosity and the other physical parameters that we
have measured, we have derived the mass accretion rate $\dot M_{\rm
acc}$ of all bona-fide PMS stars. The median value of $\dot M_{\rm acc}$
is $3.9 \times 10^{-8}$\,\Msolar\,yr$^{-1}$. This value is about 50\,\%
higher than that measured in Paper\,I for a population of 133 PMS stars 
in the field of SN\,1987A, owing to the much younger median age of PMS
objects in NGC\,346.  

\item
Thanks to the unprecedented size of our PMS sample and of its spread in
mass and age, we have been able to study the evolution of the mass
accretion rate as a function of stellar parameters. Regardless of the
mass of the star, our analysis shows that the mass accretion rate
decreases with roughly the square root of the age, or about three times
slower than predicted by current models of viscous disc evolution, and
that more massive stars have systematically higher mass accretion rate
in proportion to their mass. A multivariate linear regression fit
reveals that $\log \dot M_{\rm acc} \simeq -0.6 \log t + \log m + c$,
where $t$ is the age of the star, $m$ its mass and $c$ a quantity that 
is higher at lower metallicity. 

\item 
The large mass accretion rates that we find imply that a considerable
amount of mass is accreted during the PMS phase, {at least in 
low-metallicity environments such as the Magellanic Clouds}. At face
value, the observed rates integrated over the entire PMS phase would
give a total accreted mass of $\sim 0.2$\,\Msolar, with negligible
dependence of the  final mass on the ZAMS, thereby possibly suggesting a
lower cut-off mass in the stellar mass function of NGC\,346. If,
however, the circumstellar discs are eroded on a time scale shorter than
the PMS lifetime, the total accreted mass will be of course smaller but
not negligible. In any case, the PMS  evolution of moderate-mass stars
($< 2$\,\Msolar) should be reconsidered and recalculated taking into
account the high $\dot M_{\rm acc}$ values, since for a given MS mass
the evolutionary time needed to reach the ZAMS will be longer than what
is currently estimated by models that assume $\dot M_{\rm acc} \equiv 0$
after the first few $10^4$\,yr. 

\end{enumerate}

\begin{acknowledgements}

We wish to thank an anonymous referee for useful comments that have
greatly helped us to improve the presentation of this work.
NP acknowledges partial support by HST-NASA grants GO-11547.06A and
GO-11653.12A, and STScI-DDRF grant D0001.82435. 

\end{acknowledgements}

\end{document}